\documentclass{article}

\usepackage {graphicx,latexsym}
\usepackage{amsmath,amsthm}
\usepackage{amsfonts}
\usepackage{amssymb}
\usepackage[utf8]{inputenc}
\usepackage{authblk}
\usepackage{mathtools}
\usepackage{xcolor}
\usepackage{braket}
\usepackage{bbold}

\usepackage{caption}
\usepackage{subcaption}

\newcommand{\nl}{\nonumber \\} 

\newcommand{\R}{{\mathbb R}}  
  
\newcommand{\Z}{{\mathbb Z}}

\DeclareMathOperator{\Tr}{Tr}

\begin{document}

\title{
Homotopy data as part of the lattice field: A first study 
}
\author{
Pietro Dall'Olio
\footnote{e-mail: \ttfamily pietro@matmor.unam.mx}
}
\author{
José A. Zapata
\footnote{e-mail: \ttfamily zapata@matmor.unam.mx}
}
\affil{
Centro de Ciencias Matemáticas, 
Universidad Nacional Autónoma de México, 
C.P. 58089, Morelia, Michoacán, México.
}

\date{}
\maketitle

\begin{abstract} 
Fields exhibit a variety of topological properties, like different topological charges, when field space in the continuum is composed by more than one topological sector. Lattice treatments usually encounter difficulties describing those properties. In this work, we show that by augmenting the usual lattice fields to include extra variables describing local topological information (more precisely, regarding homotopy), the topology of the space of fields in the continuum is faithfully reproduced in the lattice. We apply this extended lattice formulation to some simple models with nontrivial topological charges, and we study their properties both analytically and via Monte Carlo simulations. 
\end{abstract}

\tableofcontents

\section{Introduction}
%

Extended lattice gauge fields were introduced in \cite{MZ1, MZ2} with a purely topological motivation; the objective was to be able to unambiguously determine a principal bundle from the gauge field (seen as a parallel transport map assigning group elements to paths which is compatible with the concatenation of paths and group multiplication), as can be done in the continuum. 
It was noticed that ordinary gauge lattice gauge fields complemented by local data describing relative homotopy classes of gauge fields (in the continuum) are enough to determine a principal bundle. 

Here we introduce extended lattice fields for sigma models. 
They are composed of a lattice field complemented by local homotopy data. The aim is exploring extended lattice fields in a simpler context. 
Additionally, 
in this article we show that 
extended fields suggest different strategies to model topological aspects of the field in numerical simulations: As we will explain, having local topological data allows us to implement a traditional Metropolis Monte Carlo algorithm in which the different topological sectors of the field are efficiently sampled.

In the case of abelian gauge fields, extended gauge fields have been implicitly considered previously, playing a prominent role in the proof of existence of a non-confining phase in compact QED with no fermions \cite{BMK77, Guth80, PPW91}. The general argument given by Polyakov \cite{Polyakov75} assumes that the path integral of the lattice model in the continuum limit, and in a semiclassical regime, is governed by the structure of the space of solutions of the gauge field in the continuum. This general argument was converted into a mathematical theorem which may be stated purely in terms of lattice fields, and without the need to reach the continuum limit \cite{BMK77, Guth80, PPW91}. This analytic  work uses Villain's periodic gaussian measure, and as explained in Section \ref{extended} (for the case of sigma models) using Villain's action, and letting its integer valued variables play a role in the space of lattice fields, is equivalent to using extended lattice fields.

We present a study exploring basic topological aspects of the field using extended fields on the lattice. We start using a simple 1-dimensional abelian system as a case study; later we define extended fields in general and present a non abelian example, and describe with some detail extended abelian fields in $2$ dimensions.

The space of fields in the continuum, 
for 
field theories 
of great interest, 
is not a connected space. 
In 
gauge theories and sigma models 
this happens frequently. In the case of gauge theories over a $d$-dimensional torus, the requirement for having a space of fields with a single connected component is that $\pi_{d-1} (G)=0$. But we know that $\pi_1 (U(n)) = \Z$ for all $n$ and that $\pi_3 (SU(n))= \pi_3 (U(n))= \Z$ for all $n >1$. 
The main reason making complicated and subtle the study of topological aspects of the field within a lattice approach is that the space of standard lattice fields is a connected space. This state of affairs leaves two possible scenarios for modeling topological properties of the field in a lattice framework: 
\begin{enumerate}
	\item 
	Excise a portion of the space of standard lattice fields, possibly of measure zero \cite{Luscher82}. 
	\item
	Complement standard lattice fields with a discrete set of extra degrees of freedom locally describing the topology of the field. These are the fields which we call extended lattice fields (ELFs). 
\end{enumerate}

Scenario (1) has been developed by the lattice community for decades, and it has tremendous achievements, and also certain shortcomings. 
Let us mention two of the shortcomings: (i) 
When working on a finite lattice, 
after a given excision in the space of lattice fields only some of the topological sectors of the field can be modeled. 
In principle, this is a problem; but 
in practice, it does not pose serious restrictions. 
(ii) A careful study shows that generically the resulting lattice field theory after an excision results in field theory violating reflection positivity \cite{GK82, Creutz04}. This result does not invalidate the resulting lattice models, since reflection positivity may be restored in the continuum limit, but it is certainly a property which cannot be overlooked. 

On the other hand, scenario (2) was proposed only recently (in the context of lattice gauge fields  \cite{MZ1, MZ2}). 
Notice that ELFs in the case of abelian lattice gauge theory are not new as mentioned at the beginning of this introduction. In this respect, it should be mentioned that calculations of the topological charge and other topological aspects of the field using ``the integer valued Villain variables'' have recently appeared \cite{AGGST19, NTU21, SG19, GGS18, SGG20, Sulej20}. 
ELFs extend these ideas to non abelian gauge groups, and while the framework needs to be further developed, it promises to be advantageous in two respects as compared with scenario (1): (i) It gives total control over topological properties of the lattice field. (ii) As mentioned above, it allows for a Monte Carlo study of topological properties of the field.

We also introduce a variant of our extended lattice fields in which the same local topological data is considered, but where the standard lattice field is restricted to take values in a discrete subset of the target space and the symmetry group is restricted accordingly to be a discrete subgroup $H\subset G$ of the symmetry group of the system. 
In the context of standard lattice gauge fields, the use of discrete subgroups of the gauge group has been thoroughly explored \cite{GK82, CJR79, Rebbi80, GK81, LM82}. The clear motivation is increasing efficiency. 
While this drastic change 
is not 
appropriate to study local aspects of the system, 
topological aspects may be well described using this severely reduced model. 
The conjecture that we begin to explore is that topological aspects of the field may be studied using our extended lattice fields restricted to a discrete subgroup and eluding the need of a continuum limit. 
We show encouraging results in this direction, but since this first study is very limited, our results are merely an invitation to perform a broader study.

The paper is organized as follows: In Section \ref{standard} we first present a simple system exhibiting nontrivial topological phenomena which will be the main case study of this article. Then we give a lattice model for the system using standard lattice fields. In Section \ref{extended} we introduce extended lattice fields for this system and give a lattice model for the system using extended fields. 
In Section \ref{DELFSection} we introduce a truncated version of extended lattice fields based in a discrete subset of the target space and a corresponding discrete subgroup of the symmetry group. 
 In Section \ref{SimulationSection} we explain how extended fields allow the implementation of a standard Metropolis Monte Carlo algorithm to sample the space of extended lattice fields, and present numerical simulations for our main example. In Section \ref{GeneralEFSection} we introduce extended fields for sigma models on base space of any dimension and with any target space. Additionally, we present a nonabelian system exhibiting nontrivial topological phenomena and give an extended lattice model describing it. The aim of the section is to show that even when our main example is very simple from several points of view, more complex systems can be modeled using extended lattice fields. Finally, in Section \ref{Summary} we come back to the two scenarios mentioned above for modeling topological aspects of fields in the lattice, and use this context to give a summary of this first study of topological aspects of fields using extended lattice fields.


\

\

\section{Our main example system and its description in terms of standard lattice fields}
\label{standard}

\

The system that we will use for this first study is a particle freely moving on a unit circle. We view it as a sigma model on a 1-dimensional spacetime with target space $U(1)$ and a global $U(1)$ symmetry. Our description of the quantum system uses euclidian time with periodic boundary conditions. The field and the space of fields are written as ${\cal M}_{cont} \ni q: [0,T] \to U(1)$. 
The space of fields in the continuum ${\cal M}_{cont} = \sqcup_n {\cal M}_{cont}(n)$ is not connected; each connected component is characterized by the winding number (topological charge) of any of its elements $q: S^1_{time} \to U(1)$. The disconnectivity of the space of fields has as consequence that the vacuum of the system may depend on an angle parameter, which is analogous to the $\theta$ parameter in QCD. 

The action and the topological charge are 
\begin{equation}\label{Scont}
	S(q) = \frac{m}{2} \int_{S^1_{time}} \dot{q}^2 dt , 
\end{equation}
\begin{equation}\label{Qcont}
	Q(q) = \frac{1}{2\pi} \int_{S^1_{time}} \dot{q} dt . 
\end{equation}
One can define a Hamiltonian operator $\hat{H}(\theta): L^2(U(1)) \to L^2(U(1))$ by $\hat{H}(\theta) = - \frac{m}{2} (\partial_\varphi - i \frac{\theta}{2\pi})^2$, where we have used $\varphi$ as angular variable for $U(1)$. Then the partition function may be defined as $Z(\theta, T) = \Tr e^{-T \hat{H}(\theta)}$. The simplicity of the system under study lets us calculate the energy spectrum, and compute the topological susceptibility exactly in the continuum.

Now we describe the system in terms of usual lattice fields. 

We write the set of $N$ vertices of the time lattice as $L^0 = \{ t_i \}$. We can number the vertices from $1$ to $N$, or label the vertices by the integres with the identification of $t_i$ and $t_{i + \lambda N}$ for any $\lambda \in \Z$. The lattice spacing and the number of vertices are related by $T = N \delta_t$. 
A lattice field is a map $q^L : L^0 \to U(1)$, 
which may also be described by the set of its evaluations $q^L = \{ q_i \in U(1) \}$. Thus, the space of lattice fields is 
\[
{\cal M}_L \simeq U(1)^N , 
\] 
and the homogeneous measure in path space is $d\mu_0 = \frac{d\varphi}{2\pi}^N$.

We will consider an action of Wilson's type: 
\begin{equation}\label{SW}
	S_W(q^L) = \sum_{ij} \frac{1}{\delta_t} (1 - {\rm Re}(D_{ij})) , 
\end{equation}
where the particle's displacement after one time step is described by $q_j= q_i D_{ij}$, and ${\rm Re}(D_{ij})$ may also be written as the cosine of the angular difference $\delta_{ij}$ defined below.

We could consider other actions. In \cite{BBCW97} an action of Manton's type, which turns out to be a classically perfect action is considered. In the next section we study in detail an action of Villain's type (or heat kernel type), and point out that it can also be considered using standard lattice fields.

Any continuous function on ${\cal M}_L$ and valued in $\Z$ is necessarily constant. 
Emulating Lüscher's construction \cite{Luscher82}, one 
introduces a nontrivial continuous topological charge function valued in $\Z$ only after the space of fields has been appropriately truncated. Then one may associate a field in the continuum to the lattice field, and calculate the charge from it. 

Consider an ``angular difference function'' associated to each link of the time lattice $\delta_{ij} : {\cal M}_L \to (-\pi , \pi]$; if the field is written as $q^L = \{ q_i = e^{i \varphi_i} \}$, the function is $\delta_{ij}(q^L) = \varphi_{j} -\varphi_i , \ {\rm mod}(2\pi)$. Clearly, $D_{ij} = e^{i \delta_{ij}}$. 
These difference functions participate in the definition of the topological charge (see (\ref{Q_L}) below). We will define a truncation such that the functions $\delta_{ij}$ are continuous when restricted to ${\cal M}_L^{MinCut}$. Then the topological charge (\ref{Q_L}) will also be continuous. 
The definition of ${\cal M}_L^{MinCut} \subset {\cal M}_L$ is 
\[
{\cal M}_L^{MinCut} = {\cal M}_L - (\cup_{ij} \delta_{ij}^{-1} (\pi)) \subset {\cal M}_L . 
\]
To get a better grasp of the space ${\cal M}_L^{MinCut}$, it is useful to consider the case of a time lattice with $N=3$ vertices which can be labeled as $t_0, t_1, t_2, t_3= t_0$. Observe that $\delta_{01} + \delta_{12} + \delta_{23}$ needs to be equal to $2\pi n$ with $n$ equal to $-1$, $0$ or $1$. Then 
\[
{\cal M}_L^{MinCut} = 
{\cal M}_L^{MinCut} (-1) \sqcup {\cal M}_L^{MinCut} (0) \sqcup {\cal M}_L^{MinCut} (1) . 
\]
We can parametrize the elements of ${\cal M}_L^{MinCut} (-1)$ using $(\varphi_0 , \delta_{01} , \delta_{12})$ as variables because $\varphi_1 = (\varphi_0 + \delta_{01}) \ {\rm mod}(2\pi)$ and $\varphi_2 = (\varphi_0 + \delta_{01}  + \delta_{12}) \ {\rm mod}(2\pi)$. Similarly, ${\cal M}_L^{MinCut} (0)$ and ${\cal M}_L^{cut} (1)$ can be parametrized using the same variables. We obtain 
${\cal M}_L^{MinCut} (-1) = 
U(1) \times (-\pi , 0) \times (-\pi , -\pi - \delta_{01})$, 
${\cal M}_L^{MinCut} (1) = 
U(1) \times (0 , \pi) \times (\pi - \delta_{01}, \pi)$, 
${\cal M}_L^{MinCut} (0) = 
U(1) \times [0 , \pi) \times (-\pi , \pi - \delta_{01}) \cup 
U(1) \times (-\pi , 0] \times (-\pi - \delta_{01}, \pi)$. 

We see that the space ${\cal M}_L^{MinCut} \subset {\cal M}_L$ has three connected components indexed by the sum of the difference operators. The space ${\cal M}_L^{MinCut}$ is topologically very different from ${\cal M}_L$, but the removed set is a measure zero set. 

There are smaller spaces where the difference operators $\{ \delta_{ij} \}$ are continuous. ${\cal M}_L^{SimpCut} \subset {\cal M}_L$ is the space where the evaluation of each of the operators $\delta_{ij}$ belong to the interval $(-\frac{\pi}{2}, \frac{\pi}{2})$. This variant of the excision of the space of fields is closer to the original proposal by Lüscher \cite{Luscher82}. This space may be used to model topological aspects of the field if the lattice has five vertices or more. 
The number of connected components of the spaces ${\cal M}_L^{MinCut}$ and ${\cal M}_L^{SimpCut}$ increases with the number of lattice links. Any number of connected components can be achieved if the number of links is sufficiently large.

In ${\cal M}_L^{MinCut}$ or ${\cal M}_L^{SimpCut}$ the ``geometrical''
topological charge may be defined as 
\begin{equation}\label{Q_L}
	Q_L(q^L) = \frac{1}{2\pi} \sum_{ij} \ \delta_{ij} . 
\end{equation}

A possible interpretation of this truncation of the space of lattice fields is that the action has been modified assuming the value $\infty$ at ${\cal M}_L - {\cal M}_L^{SimpCut}$. This is one of the motivations behind \cite{BGPW10}, which explores describing truncations of the space of lattice fields, by abrupt changes in the action, which allow the authors to model a desired set of topological sectors of the field. 

Now the partition function may be defined as 
\[
Z_L(\theta, T) = \int_{{\cal M}_L^{SimpCut}} d\mu_0 \ e^{- (S_W(q^L) + i \theta Q_L(q^L))} ,
\] 
where $d\mu_0 = \frac{d\varphi}{2\pi}^N$.   

A topological observable is the topological susceptibility 
\[
\chi_L (T, \theta)= \frac{\langle Q_L^2 \rangle^L_{\theta, T}}{T} . 
\]

An observable containing information about the energy gap is the correlation function 
\[
\langle \bar{q}_0 \ q_l \rangle^L_{\theta, T} . 
\]

We report on simulations that measure these two observables in Section \ref{SimulationSection}.

\section{Description in terms of extended lattice fields}
\label{extended}

In this section we introduce extended lattice fields for the system described at the beginning of the previous section. Extended lattice fields for general sigma models are defined in Section \ref{GeneralEFSection}, and they are the natural analog of the extended lattice gauge fields introduced in \cite{MZ1, MZ2}.

The main difference between standard lattice fields and extended lattice fields is that extended lattice fields assign variables to lattice vertices and also to lattice links (and in the general case described in Section \ref{GeneralEFSection}, variables are assigned to cells of every dimension). 
The new information contained in the field captures the homotopy type of possible fields in the continuum which are compatible with lattice data at lower dimensional cells; see the example below.

The set of links of the time lattice is $L^1 = \{ [t_i , t_j] \}_{ij \ neighbors}$.  
An extended lattice field is composed of a pair of maps $q^{EL} = (q^0, q^1) \in {\cal M}_{EL}$ where $q^0 = q^L: L^0 \to G$ is the standard lattice field, 
and $q^1 : L^1 \to \mbox{Ext}(q^0)$; the field may be described by the variables 
$(\{ q_i \in U(1) \}_i , \{ [q|_{[t_i, t_j]}] \in \mbox{Ext}(q_i , q_j)\}_{ij})$. Here $\mbox{Ext}(q_i , q_j)$ denotes the set of homotopy classes of curves in $G$ starting at $q_i$ and finishing at $q_j$. If the target space $G$ is simply connected, there is no information in $q^1$. In the case of $G=U(1)$, once $q^0$ has been evaluated, we can set a correspondence between the set of possible evaluations of $q^1$ and $\Z^N$ (one copy of the integers for each lattice link). This correspondence, however, is not canonical. In the Appendix we describe the $1$-dimensional $U(1)$-extended lattice field as a pair composed of a standard $U(1)$ field and a standard $\Z$-lattice gauge field. We wonder if this realization of the extended lattice field could serve as a point of coincidence with \cite{AGGST19, NTU21, SG19, GGS18, SGG20, Sulej20}.

A local description of the space of extended lattice fields may be given by assigning a triple of variables $(q_i, q_j; [q|_{[t_i, t_j]}])$ to each lattice link $[t_i, t_j]$; then the space of extended lattice fields restricted to $[t_i, t_j]$ is 
\[
{\cal M}_{EL}|_{[t_i, t_j]} = U(1)^2 \times \mbox{Ext}(q_i , q_j) , 
\]
which is the analog of the space of extended lattice gauge fields introduced in \cite{MZ1, MZ2}. 
Below we give a more concrete description, which will let us describe the complete space ${\cal M}_{EL}$. 

\begin{figure}
\includegraphics[width=0.4\linewidth]{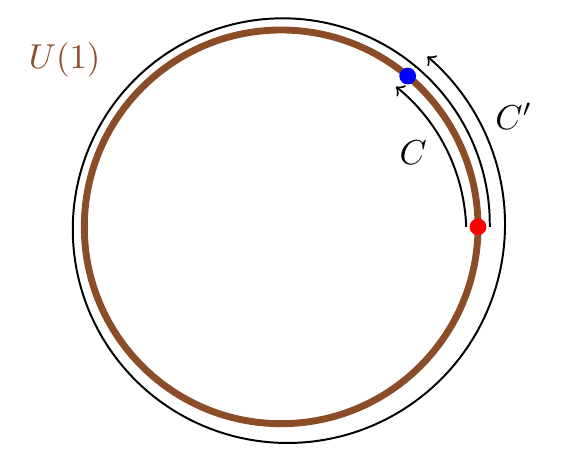}
\caption{Two curves in $U(1)$ with the same endpoints which cannot be deformed to each other.}
\label{curves}
\end{figure}

A set of relative variables simplifies the description of extended fields. 
Let us begin the description of this idea in the context of standard lattice fields. 
In terms of standard lattice fields, we could give the position of the particle at the initial time $q_0$ and also give the $U(1)$-displacement $D_{01}$ to know $q_1 = q_0 D_{01}$ (and in general $q_k = q_0 \prod D_{ij}$, where the product runs over all the links in a segment from $t_0$ to $t_k$). Thus, we could specify the standard lattice field in terms of $( q_0; \{ D_{ij} \})_L$. 
Now let us turn to the context of extended lattice fields. Given the position $q_0$ at the initial time, the extended lattice field restricted to $[t_0, t_1]$ may be determined by including a real number describing the relative position $(q_0 \in U(1), x_{01} \in \R)$. As mentioned above, we can calculate $q_1 = q_0 e^{i x_{01}} \in U(1)$. Now notice that $(q_0 , x'_{01})$ with $x'_{01} = x_{01} + 2\pi n$ leads to the same $q_1$ as $(q_0 \in U(1), x_{01})$, but if $n\neq 0$ they represent different extended lattice fields restricted to $[t_0, t_1]$ (see Fig. \ref{curves}). 

Thus, the alternative description of ${\cal M}_{EL}|_{[t_i, t_j]}$ is simply ${\cal M}_{EL}|_{[t_i, t_j]} = U(1) \times \R$. 
The reason for the space of ``relative position variables'' being the real numbers is that the ``universal covering group'' of $U(1)$ is $\R$. 
In Section \ref{GeneralEFSection} we will give a definition of extended lattice fields for general sigma models. We will also see the example of a 1-dimensional model representing rigid body motion with $G = SO(3)$; in that case, the universal covering group where the ``relative position variables'' live is $SU(2)$, which covers $SO(3)$ twice. 

We see that a description of an extended lattice field using this variables is $(q_0 \in U(1); \{ x_{ij} \in \R \})_{EL}$, where the periodic boundary conditions demand that $\sum x_{ij} = n 2\pi$ for some $n \in \Z$. Thus, the space of extended lattice fields is 
\begin{equation}\label{SpaceOfELf}
	{\cal M}_{EL} = \sqcup_n \ {\cal M}_{EL}(n) , \quad \mbox{ with } \quad 
	{\cal M}_{EL}(n) = U(1) \times \R^{N-1} . 
\end{equation}

The space of extended lattice fields is a covering space of the space of standard lattice fields $Pr: {\cal M}_{EL} \to {\cal M}_L$. The easiest way to see it is describing an extended field as a pair of maps $q^{EL} = (q^0, q^1)$; $Pr$ simply forgets the $q^1$ map and sets $q^L = Pr(q^{EL}) = q^0$. In terms of relative variables, we get $Pr ((q_0 \in U(1); \{ x_{ij} \in \R \})_{EL}) = (q_0 \in U(1); \{ e^{i x_{ij}} \in U(1) \})_L$. 
Understanding the fibers $Pr^{-1}(q^L)$ of the covering map would let us prescribe an extended lattice gauge field in terms of a pair $(q^L, z)$, where the first component chooses the fiber, and the second specifies a point in the fiber. In the Appendix, we show that $z$, the second component of the pair, is related to a $\Z$-gauge field or a $\Z$-sigma model living on the dual lattice. When restricted to a link $(ij)$, the nature of the relation between the variables $(q^L = (q_i, q_j), z_{ij})$ and the variables $(q_i , x_{ij})$ is analogous to the relation between an infinite collection of circles $(q_j, z_{ij}) \in U(1) \times \Z$ and the real line $x_{ij} \in \R$; in order to match them, one has to cut each of the circles and paste the resulting intervals together making an infinite line. 

In lattice field theory it is crucial to have a homogeneous measure in the space of fields. In ${\cal M}_L = U(1)^N$ the measure is the Haar measure $d\mu^L_{Haar} = \frac{d\varphi}{2\pi}^N$, which is a probability measure. The space ${\cal M}_{EL}$ is not compact; it has a homogeneous measure, but it is not normalizable. When using variables $(q_0 \in U(1); \{ x_{ij} \in \R \})_{EL}$, the Haar measure in ${\cal M}_{EL}$ is written as $d\mu^{EL}_{Haar} = \frac{d\varphi_0}{2\pi} \ (\frac{dx}{2\pi})^{N-1}$. On the other hand, if we use variables $(\{ q_i \in U(1) \}_i , \{ [q|_{[t_i, t_j]}] \in \mbox{Ext}(q_i , q_j)\}_{ij})$, the Haar measure is written as $d\mu^{EL}_{Haar} = \frac{d\varphi}{2\pi}^N \  \mu_{count}^N$, where $\mu_{count}$ is the counting measure in the space $\mbox{Ext}(q_i , q_j) \simeq \Z$ (where, we remind the reader that the parametrization of $\mbox{Ext}(q_i , q_j)$ with $\Z$ is not canonical). 
It is easy to see that if $U \subset {\cal M}_{EL}$ is a subset such that $Pr |_U$ is invertible, then the measure $(Pr |_U)_\ast d\mu^{EL}_{Haar}$ in $Pr (U)$ is equal to $d\mu^L_{Haar}$. 

Notice that ${\cal M}_{EL}$ is topologically very different from ${\cal M}_L$; it is a union of an infinite countable number of connected components, and each connected component is non compact. The space of fields in the continuum ${\cal M}_{cont}$ is also a union of an infinite countable number of non compact connected components. 

The action and the topological charge are\footnote{The units of the lattice spacing $\delta_t$ are determined by fixing the inertial moment to unity.} 
\begin{equation}\label{S_EL}
	S_{EL}(q^{EL}) = \sum_{ij} \frac{1}{2}  \left(\frac{x_{ij}}{\delta_t} \right)^2  \delta_t , 
\end{equation}
\begin{equation}\label{Q_EL}
	Q_{EL}(q^{EL}) = \frac{1}{2\pi} \sum_{ij}  \left(\frac{x_{ij}}{\delta_t} \right)  \delta_t . 
\end{equation}

A field $q$ in the continuum induces an extended lattice field $q^{EL}= {\rm Ev}_{EL} (q)$. An obvious, but important property of \eqref{Q_EL} is that for any field $q$ we have $Q_{EL}({\rm Ev}_{EL} (q)) = Q(q)$. 

The partition function written in terms of extended fields is 
\begin{equation}\label{Z_EL}
	Z_{EL}(\theta, T) = \int_{{\cal M}_{EL}} d\mu^{EL}_{Haar} \ e^{- (S_{EL}(q^{EL}) + i \theta Q_{EL}(q^{EL}))} . 
\end{equation}

Notice that, when restricted to a topological sector ${\cal M}_{EL}(n)$, the physical measure shown in the path integral is proportional to a gaussian measure (shifted by the topological term). The relative weight of the different topological sectors is discused further ahead in this section. 

The transfer matrix, an operator in $L^2(U(1))$ with matrix elements $\braket{\varphi_j | \hat{T}(\theta, \delta_t) | \varphi_i}$, is simple to write 
\begin{eqnarray}\label{T_EL}
	\frac{1}{2\pi}  
\int_{\R} dx_{ij} \sum_n  \ \delta \left(\varphi_j + 2\pi n , \varphi_i + x_{ij} \right)_{\R} 
e^{- \frac{1}{2 \delta_t}
x_{ij}^2 - i \theta \frac{x_{ij}}{2\pi}} \nonumber \\
= \frac{1}{2\pi}  \sum_n  
e^{- \frac{1}{2 \delta_t}
(\varphi_j - \varphi_i + 2\pi n )^2 - i \theta \frac{\varphi_j - \varphi_i + 2\pi n }{2\pi}} , 
\end{eqnarray}
which, apart from the $\theta$ term, is recognizable as the transfer matrix corresponding to the Villain action for the standard lattice field. Indeed, when the gaussian measure in ${\cal M}_{EL}$ is pushed forwards with $Pr$ to ${\cal M}_L$ we get the ``periodic gaussian'' measure or Villain measure. 
The physical interpretation of the push-forward operation $Pr_\ast$ is clear; we are integrating over the degrees of freedom not present in standard lattice fields.

We know that due to the rotational $U(1)$ symmetry, in the angular momentum basis the transfer matrix $\braket{ k_j| \hat{T}(\theta, \delta_t) |k_i }$ is diagonal. In order to do the calculation, we first rewrite the integrations, sums and deltas involved. The main step is to interchange a delta function in $\R$ by a product of a delta function in $S^1$ and a Kronecker delta 
\begin{eqnarray}
&&\frac{1}{2\pi} \int d\varphi_i \ \frac{1}{2\pi} \int d\varphi_j \
\frac{1}{2\pi}  
\int_{\R} dx_{ij} \sum_n \ \delta \left(\varphi_j + 2\pi n , \varphi_i + x_{ij} \right)_{\R}  \\
&&= \frac{1}{2\pi} \int d\varphi_i \ \frac{1}{2\pi} \int_{\R} dx_{ij}  
\ \sum_n  \ \frac{1}{2\pi} \int d\varphi_j 
\ \delta_{2\pi n, \varphi_i - \varphi_j + x_{ij}} 
\ \delta \left(\varphi_j , \varphi_i + x_{ij} {\rm mod} 2\pi \right)_{S^1} . \nonumber 
\end{eqnarray}
Then $\braket{ k_j| \hat{T}(\theta, \delta_t) |k_i }$ evaluates to 
\begin{eqnarray}\label{T_ELdiagonalized}
&&\frac{1}{2\pi} \int d\varphi_j \ \frac{1}{2\pi} \int d\varphi_i \
e^{-i k_j \varphi_j} e^{i k_i \varphi_i}
\braket{ \varphi_j| \hat{T}(\theta, \delta_t) |\varphi_i } = \nonumber \\
&&= \delta_{k_j , k_i} \ \frac{1}{2\pi} \int dx \
e^{-ik_i x} e^{-\frac{1}{2 \delta_t} x^2 - i \theta \frac{x}{2\pi}} \nonumber \\
&&=  \delta_{k_i , k_j} \ \lambda \ e^{-\frac{\delta_t}{2}
(k_i + \frac{\theta}{2\pi})^2} , 
\end{eqnarray}
where in the first step we have started from the first line of equation (\ref{T_EL}), and used the previous manipulation of integrals and delta functions. The numerical factor $\lambda$ is independent of $k_i$.

We are ready to investigate physical observables. Here we will study the topological susceptibility and the correlation function. After the calculations are finished, it will be interesting to notice that both observables have well defined thermodynamic limits.

We could proceed to evaluate the topological susceptibility as we would evaluate any expectation value, but it is more instructive to see that the evaluation of the topological charge probability distribution  $P(Q)$ (calculated at $\theta = 0$) has a simple formula 
\begin{eqnarray}
	P(Q) \propto & \! \! \int_{{\cal M}_{EL}(n = Q)} d\mu^{EL}_{Haar} \ e^{- S(q^{EL})} \nl
	\propto &  \! \! \exp( - \frac{1}{2} (\frac{(2\pi Q)^2}{T})) , 
\end{eqnarray}
where the proportionality factors are independent of $Q$. 
A simple proof of this formula follows from considering a parametrization of the space ${\cal M}_{EL}(n=Q)$ in terms of a configuration for which all $x_{ij}$ are equal to $2\pi Q/T$ and local perturbations which do not change the topological charge. The Boltzmann weight factorizes as a product of a factor corresponding to the uniform configuration (which comes out of the integral), a factor which is linear in the perturbation (which does not contribute to the integral due to parity in the space of perturbations), and a quadratic factor (which is independent of $Q$). 
From this probability distribution the calculation of the susceptibility follows directly 
\begin{equation}\label{susceqEL}
	\chi_{EL}(T) = \frac{1}{T}
	\frac{\sum_{Q  \in \mathbb{Z}}Q^2 P(Q)}{\sum_{Q  \in \mathbb{Z}}P(Q)} = \frac{1}{T} \frac{\sum_{Q \in \mathbb{Z}} Q^2 e^{-2\pi^2  Q^2/T}}{\sum_{Q \in \mathbb{Z}} e^{-2\pi^2  Q^2/T}} , 
\end{equation}
which, in the thermodynamic limit goes to $1/(4 \pi^2)$. See Figure \ref{susctop}. 

A view of our results from the perspective of standard lattice fields is illuminating: 
Each extended lattice field has a topological charge. On the other hand, given an ordinary lattice field $q^L$, the set of its preimages $P^{-1}(q^L) \subset {\cal M}_{EL}$ is a discrete set whose elements have all possible topological charges. A related observation was made in \cite{BBCW97}.  

Now we calculate the correlation function 
\begin{equation}\label{correqEL}
\langle \bar{q}_0 q_l \rangle_{\theta, T}^{EL} =
\frac{1}{Z_{EL}(\theta, T)}
 \Tr(\hat{q}^\dagger \hat{T}^l \hat{q} \hat{T}^{N-l})	 =  \frac{e^{-\frac{t(T -t)}{2 T}}  \displaystyle\sum_{k \in \mathbb{Z}} e^{-\frac{T}{2 }\left(k - \frac{\theta}{2\pi} -\frac{t}{T} \right)^2}}{ \displaystyle\sum_{k \in \mathbb{Z}} e^{-\frac{T}{2}\left(k - \frac{\theta}{2\pi} \right)^2}}. 
\end{equation}
See Figure \ref{corrfig}. In the thermodynamic limit, in which $T$ goes to infinity while physical time $t = l \delta_t$ remains fixed, the correlation function goes to $e^{-l \delta_t/2}$.

Notice that our calculations of the topological susceptibility and the correlation function (of points separated by a given physical time $t = l \delta_t$) are independent of the lattice spacing $\delta_t$; in fact, the calculations produce the same result as in the continuum. The energy spectrum, which may be calculated from the Hamiltonian associated to the transfer matrix (\ref{T_ELdiagonalized}), is also independent of the lattice spacing. The reason behind this fact is that our model has a quantum perfect action. We will comment a bit more on this subject below.

\section{Description in terms of discrete extended lattice fields}
\label{DELFSection}

In math there is a hierarchy in which sets come first, then they are dressed with topology, and then they are dressed with a differential structure (one can continue giving them extra structure like a metric structure). In this hierarchy each higher level requires more complication. A natural question arises: Can the description of topological aspects of the field be isolated and simplified? 

In this section we will give an affirmative answer to that question. We will restrict the possible evaluations of the lattice field to a discrete subgroup $H \subset G$, but keep the homotopy information. 
We are proposing to use this simplified framework for modeling topological (global) aspects of the field only, and not for aspects of the field which are local in nature. 
The promised simplification is the possibility of eluding the need of a further continuum limit in which the subgroup becomes finer.
Our results are encouraging, but clearly further research would be necessary to corroborate or discard the conjectured simplification. 
It should be mentioned that the restriction of gauge fields in the lattice to have values in discrete subgroups of the gauge group was extensively studied in the 70s and 80s, see for example \cite{GK82, CJR79, Rebbi80, GK81, LM82}. 
Our proposal is different in two respects: First, our fields include homotopy data (regarding $G$). Second, we propose to study global observables only. 
In Section \ref{GeneralEFSection} we numerically study the case $G=SO(3)$ with encouraging results. 


Discrete extended lattice fields for sigma models in the general case are briefly described in Section \ref{GeneralEFSection}. In this section we introduce them in the case of a 1-dimensional base and for the discrete subgroups $\Z_n$ realized as subgroups of $U(1)$. 
We will write $e^{i 2\pi m/n} \in \Z_n \subset U(1)$. 
A discrete extended lattice field in a time lattice based on $\Z_n \subset U(1)$ is composed of a pair of maps $q^{DEL} = (q^0, q^1) \in {\cal M}_{DEL}$ where $q^0 : L_0 \to \Z_n$ is the standard lattice field but restricted to take values in $\Z_n$, and $q^1 : L_1 \to \mbox{Ext}(q^0)$. 
The field may be described by the variables 
$(\{ q_i \in \Z_n \}_i , \{ [q|_{[t_i, t_j]}] \in \mbox{Ext}(q_i , q_j)\}_{ij})$, where $\mbox{Ext}(q_i , q_j)$ denotes the set of homotopy classes of curves in $U(1)$ starting at $q_i$ and finishing at $q_j$. 
We restrict the allowed evaluations of the standard lattice field to $\Z_n$, but the topological information stored in the map $q^1$ is the homotopy class of a map from $[t_i, t_j]$ to $U(1)$. 

The space of discrete extended lattice fields restricted to $[t_i, t_j]$ is 
${\cal M}_{DEL}|_{[t_i, t_j]} = \Z_n^2 \times \mbox{Ext}(q_i , q_j)$. 
Clearly, this space of fields is just a discrete set. Let us compare it with ${\cal M}_{EL}|_{[t_i, t_j]}$. The factor $\mbox{Ext}(q_i , q_j)$ is exactly the same, all the difference is that the factor $U(1)^2$ in ${\cal M}_{EL}|_{[t_i, t_j]}$ has been replaced with its discrete analog $\Z_n^2$. It is known that discrete subgroups of Lie groups $H \subset G$ can be associated to the $0$-skeleton of a cellular decomposition of $G$ which are invariant under the right and left actions of $H$ on $G$ (see e.g. \cite{Coxeter54}). In this case, $\Z_n$ gives the $n$ vertices of a cellular decomposition of $U(1)$ which also has $n$ links. 
Since this is true for each time subinterval $[t_i, t_j]$, even when ${\cal M}_{DEL}$ is a discrete set, it has the interpretation of being the 0-skeleton of a cellular decomposition of ${\cal M}_{EL}$.

We introduce relative variables simplifying the description. We give the position of the particle at the initial time $q_0 \in \Z_n$ and a relative position variable $x_{ij} \in \frac{2 \pi}{n} \Z \subset \R$ for each time interval $[t_i, t_j]$ (restricted by the periodic boundary conditions).

The prescriptions for the topological charge and the action for our system of interest are obtained simply by evaluating $S_{EL}$ and $Q_{EL}$ as given in equations \eqref{S_EL}, \eqref{Q_EL} in $q^{DEL} \in {\cal M}_{DEL} \subset {\cal M}_{EL}$. For the partition function we obtain  
\begin{equation}\label{Z_DEL}
	Z_{DEL}(\theta, T) = \sum_{{\cal M}_{DEL}}  \ e^{- (S_{EL}(q^{DEL}) + i \theta Q_{EL}(q^{DEL}))} , 
\end{equation}
where the role played by the Haar measure in ${\cal M}_{EL}$ is played by the counting measure in ${\cal M}_{DEL}$. 
This is a good opportunity to get a better grasp of the relation between using fields valued in $U(1)$ versus using fields restricted to be valued in $\Z_n \subset U(1)$. 
There are exactly $n$ irreducible complex representations of $\Z_n$ which are all 1-dimensional $\{ R_m(e^{i 2\pi k/n}) = e^{i 2\pi mk/n} \}_{m \in \Z_n}$. They give a basis for complex valued functions on $\Z_n$; a very good hands on review can be found in \cite{CreutzBook}. 
We evaluate the elements of the ordinary Fourier basis $\{ \psi_k (\varphi) = e^{ik \varphi} \}_{k \in \Z}$ on fields valued in $\Z_n$, and then expand the resulting functions in terms of the basis $\{ R_m \}_{m \in \Z_n}$. We obtain $\psi_k|_{\Z_n} = R_{[k] {\rm mod} n}$. Thus, the interpretation of working with fields valued in $\Z_n$ is not to truncate the energy spectrum to make the space of quantum states $n$-dimensional; instead, the $\Z_n$ theory reduces the dimension of the space of quantum states by identifying quantum states with vastly different energies. A concrete realization follows from the calculation of the transfer matrix $\braket{\varphi_j | \hat{T}_{DEL}(\theta, \delta_t) | \varphi_i}$, which up to normalization, may be obtained by evaluating $\hat{T}_{EL}(\theta, \delta_t)$ on angles in $\Z_n$. Due to the global $\Z_n$ symmetry of the action and the topological charge, it is diagonal in the $\{ R_m \}_{m \in \Z_n}$ basis. We obtain 
\begin{equation}\label{T_DELdiagonalized}
\braket{ m_j| \hat{T}_{DEL}(\theta, \delta_t) |m_i }
=  \delta_{[m_i] , [m_j]} \ 
\Lambda_{m_i}(\theta, \delta_t) , 
\end{equation}
where $\Lambda_{m_i}(\theta, \delta_t) = \lambda \sum_{k \in [m_i]} e^{-\frac{\delta_t}{2}
(k_i + \frac{\theta}{2\pi})^2}$ and $\lambda$ is a numerical factor which is independent of $m_i$. From this expression we can calculate $\ln(\Lambda_m / \Lambda_{m'})$; the fact that the result does not depend linearly on $\delta_t$ tells us that the model is not quantum perfect. We can also see that as $n$ becomes larger we get a larger space of states, and each $\Lambda_m$ becomes more and more dominated by the smallest eigenvalue in the class $[m]$ resulting in a better fit of the $\mathbb{Z}_n$ model to the $U(1)$ model. 
Notice that a continuum limit, in which the discrete subgroup becomes finer is impossible for most groups.

We can evaluate the topological susceptibility analytically. The more instructive way to do it is to directly evaluate the topological charge probability distribution $P_{DEL}(Q, N, n)$ (calculated at $\theta = 0$). Recall that the total time has been subdivided into $N$ subintervals $T = \delta_t N$; if it happens that $\frac{N}{n} \in \Z$, then the motions of ``uniform angular velocity in $\Z_n$'' are allowed by the periodic boundary conditions. Those histories are the extrema of the action governing our gaussian distribution. In this case, it is easy to verify that 
\begin{equation}
P_{DEL}(Q, N, n) \propto \exp( - \frac{1}{2} (\frac{(2\pi Q)^2}{T})) , 
\end{equation}
which is the same relation followed by $P_{EL}(Q)$. Thus, in those cases the topological susceptibility calculated using fields valued in $\Z_n$ agrees with the topological susceptibility calculated using fields valued in $U(1)$. 

On the other hand, given $N$ a number of time subintervals there are values of $n$ such that $\frac{N}{n}$ is not an integer, but such that there is $l < n$ such that $\frac{N}{nl}$ is an integer. This means that using fields valued in $\Z_n$ the topological sectors of the field associated to topological charges which are multiples of $l$ have histories which extremize the action, while the other sectors do not. For those cases $P_{DEL}(Q, N, n)$ does not follow the usual distribution.

The correlation function can be computed from the transfer matrix \eqref{T_DELdiagonalized} 
\begin{equation}\label{correqDEL}
\langle \bar{q}_0 q_l \rangle_{DEL, \theta, T} =
\frac{1}{Z_{DEL}(\theta, T)}
 \Tr(\hat{q}^\dagger \hat{T}_{DEL}^l \hat{q} \hat{T}_{DEL}^{N-l}). 
\end{equation}
As mentioned earlier, for functions that are not global in character, we cannot expect to reproduce the results obtained using using extended lattice fields. 
See Figure \ref{corr_discr} and the comment in the last paragraph of Section \ref{SimulationSection}.

\section{Simulations of topological phenomena using extended lattice fields}
\label{SimulationSection}

Although the 1-dimensional example described above is analytically solvable, it is interesting to study it also via Monte Carlo simulations%
\footnote{ 
One reason for the interest is that it would be simple to deform our integrable model for the free particle on the circle using a periodic potential which makes it not solvable. We do not study such deformations, but in Section \ref{DELFSection} we mentioned a versions of our model truncated by restricting the variables to discrete subgroups, in which case the resulting models are not quantum perfect. We report simulations for those models in this section. 
}. 
The novel perspective offered by the use of extended fields suggests different numerical strategies to capture topological properties of the field. 

It is often claimed \cite{BFG15, BWD15} that a direct evaluation of the topological susceptibility via local update algorithms is unfeasible (due to the very low frequency of shifts among different topological sectors), and that hence cluster algorithms \cite{W89} are required. 
Additionally, if certain reasonable conditions hold, there are clever indirect ways of evaluating the topological susceptibility which would be valid even if one had access to a single topological sector \cite{BFG15, BWD15} of field space.
On the other hand, the extended lattice framework gives a complete control on topological aspects of the field. We could restrict to any given topological sector, and investigate only that sector. More importantly, we can generate a Markov chain in field space using the traditional Metropolis algorithm \cite{MU49} without getting stuck in a single topological sector. 

We perform simulations to study a quantum particle moving on a circle using a time lattice with extended lattice fields using variables $x_i \equiv x_{i, i+i} \in \mathbb{R}$ associated to the time intervals $[t_i, t_{i+1}]$ as described in Section \ref{extended}. 
We evaluate the topological susceptibility \eqref{susceqEL} by randomly updating the extended variables as explained below: 
For a given run of the simulation, we select a number $1 < l \leq N$. At each iteration, we randomly choose an integer $\delta_n \in \{-1, 0, 1\}$ and a subset of $l$ adjacent variables $\{ x_i, x_{i+1} \dots x_{i+l-1} \}$. A local potential variation changing those variables is generated; ideally, the potential variation satisfies 
$\sum_{k=i}^{i+l-1} \delta x_k \ = \ 2\pi \delta_n$, but it is random otherwise. 
In practice, we write $\delta x_k = \delta^\varphi_k + \delta^n_k$ with $\delta^\varphi_k \in [-\epsilon , \epsilon ]$, and require that the set of these small numbers satisfies 
$\sum_{k=i}^{i+l-1} \delta^\varphi_k \ = \ 0$, but are random otherwise; on the other hand, we declare $\delta^n_k = 2 \pi \delta_n / l$. 
In this way, the potential change in topological charge caused by the potential variation of the field is 
\begin{equation}
\delta Q = \frac{1}{2\pi}\sum_{j=0}^{N-1} \delta x_j = \delta n .
\end{equation}

The action's variation caused by an update $\{ \delta x_i \}$ of a typical configuration $\{ x_i \}$ is given by
\begin{equation}
\label{deltaction}
\delta S = \frac{1}{\delta_t} \sum_{j =i}^{i+l-1} \delta x_i \left( x_i + \frac{\delta x_i}{2} \right) . 
\end{equation}
Notice that if the lattice spacing $\delta_t$ is too small, i.e. for fine lattices, its magnitude will be large for the updates that shift the topological charge ($|\delta^n_i| = \frac{2\pi}{l}$) causing the rejection of most of these configurations by the Metropolis algorithm.
In the case of a model with a quantum perfect action like this one, the problem is easily overcome by using large values of $\delta_t$ (a more universally useful solution is discussed in the next paragraph). 
The topological susceptibility has been evaluated for a lattice of size $T= N \delta_t = 2000$ (in arbitrary units) for different values of the lattice spacing (see Table \ref{t1}). The results are all compatible, within statistical errors, with the analytical thermodynamic limit. It is remarkable that the correct result is generated also in the case with only two lattice nodes (last entry in the table). All the simulations have been performed by implementing the Metropolis algorithm with $5\times 10^6$ iterations, the first $10^3$ of which used for thermalization,
registering the topological susceptibility each $10$ iterations, a step size heuristically fixed by checking the reproducibility of the results over several simulations. 
\begin{table}[!h]
\centering
\begin{tabular}{|c|c|c|c|}
	\hline
	 $\delta_t$ & $\chi_t = \langle Q^2 \rangle/T\,\,\, (T = 2000)$ \\
	\hline\hline
	5 & 0.0252(7)\\
	\hline
	8 & 0.0252(8)\\
	\hline
	10 & 0.0253(5)\\
	\hline
	20 & 0.0252(2)\\
	\hline
	25 & 0.0252(1)\\
	\hline
	40 & 0.0253(8)\\
	\hline
	100 & 0.0253(1)\\
	\hline
	200 & 0.0253(6)\\
	\hline
	1000 & 0.0252(5) \\
	\hline
\end{tabular}
\caption{Topological susceptibility $\chi_t$ at fixed size $T=2000$ for different lattice spacings $\delta_t$. The values are independent of the lattice spacing and 
compatible with the analytical value in the thermodynamic limit $\chi_t = \frac{1}{4\pi^2} \simeq 0.02533 $.}
\label{t1}
\end{table}

\begin{figure}
\includegraphics[width=1.0\linewidth]{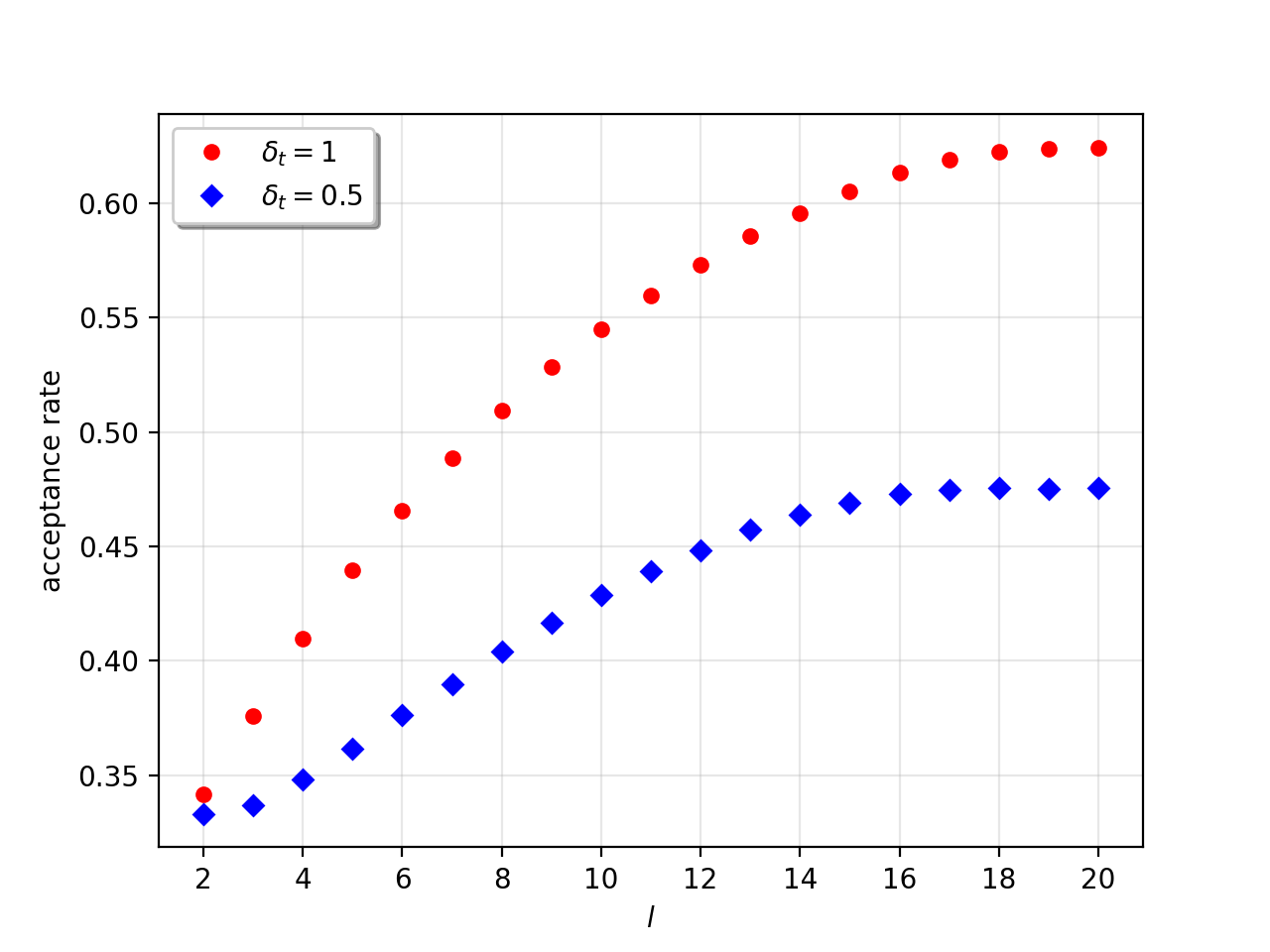}
\caption{Increasing of the acceptance rate due to the increasing of the number $l$ of neighboring variables $x$ involved in each update. The two plots correspond to two different values of the lattice spacing $\delta_t$.}
    \label{accrate}
\end{figure}

Now we discuss a strategy to obtain simulations with higher acceptance rate that applies to any abelian theory and does not rely on having a quantum perfect action. A closer look at the expected value of the change of the action due to an update of a typical field, considering that $\langle \delta x_i \rangle = 0$, is $\langle \delta S \rangle \sim \frac{l}{\delta_t} \langle (\delta^n_j )^2\rangle \sim \frac{(2\pi)^2}{l \delta_t}$. This suggests that a way to counterbalance the barrier resulting from a small lattice spacing is to increase the number $l$ of variables among which the topological shift is distributed (see fig. \ref{accrate}). The improvement of the values of the susceptibility due to the increasing of the number $l$ of variables involved in each update, is shown in Table \ref{t2}. 

\begin{table}[!h]
\centering
\begin{tabular}{|c|c|c|c|c|}
\hline
$ N $ & $\delta_t$ & $l$ & $ \chi_t$ & $\chi^{analytic}_t$ \\
\hline \hline
 &  & 2 & 0.0266(9) & \\
20 & 1 & 10 & 0.0253(5) & 0.02528 \\
&  & 20 & 0.0252(8) & \\
\hline
 &  & 2 & 0.0174(6) & \\
20 & 0.5 & 10 & 0.0218(3) & 0.02196 \\
 & & 20 & 0.0219(6) & \\
\hline
 &  & 2 & 0.1900(0) & \\
20 & 0.25 & 10 & 0.0077(6) & 0.00743 \\
 &  & 20 & 0.0074(1) & \\
\hline
 &  & 2 & 0.0028(2) & \\
40 & 0.25 & 10 & 0.0218(8) & 0.02196 \\
 &  & 20 & 0.0220(6) & \\
\hline
 &  & 10 & 0.0084(0) & \\
40 & 0.125 & 20 & 0.0078(4) & 0.00743 \\
 &  & 40 & 0.0075(9) & \\
\hline
\end{tabular}
\caption{Topological susceptibility $\chi_t$ for lattices with $N=20$ and $N=40$ lattice nodes. For smaller values of the lattice spacing $\delta_t$ the analytic results $\chi^{analytic}_t$ (eq. \eqref{susceqEL}) are best approximated by the simulations that 
use a larger number $l$ of variables in each update.}
\label{t2}
\end{table}

\begin{figure}
\includegraphics[width=1.0\linewidth]{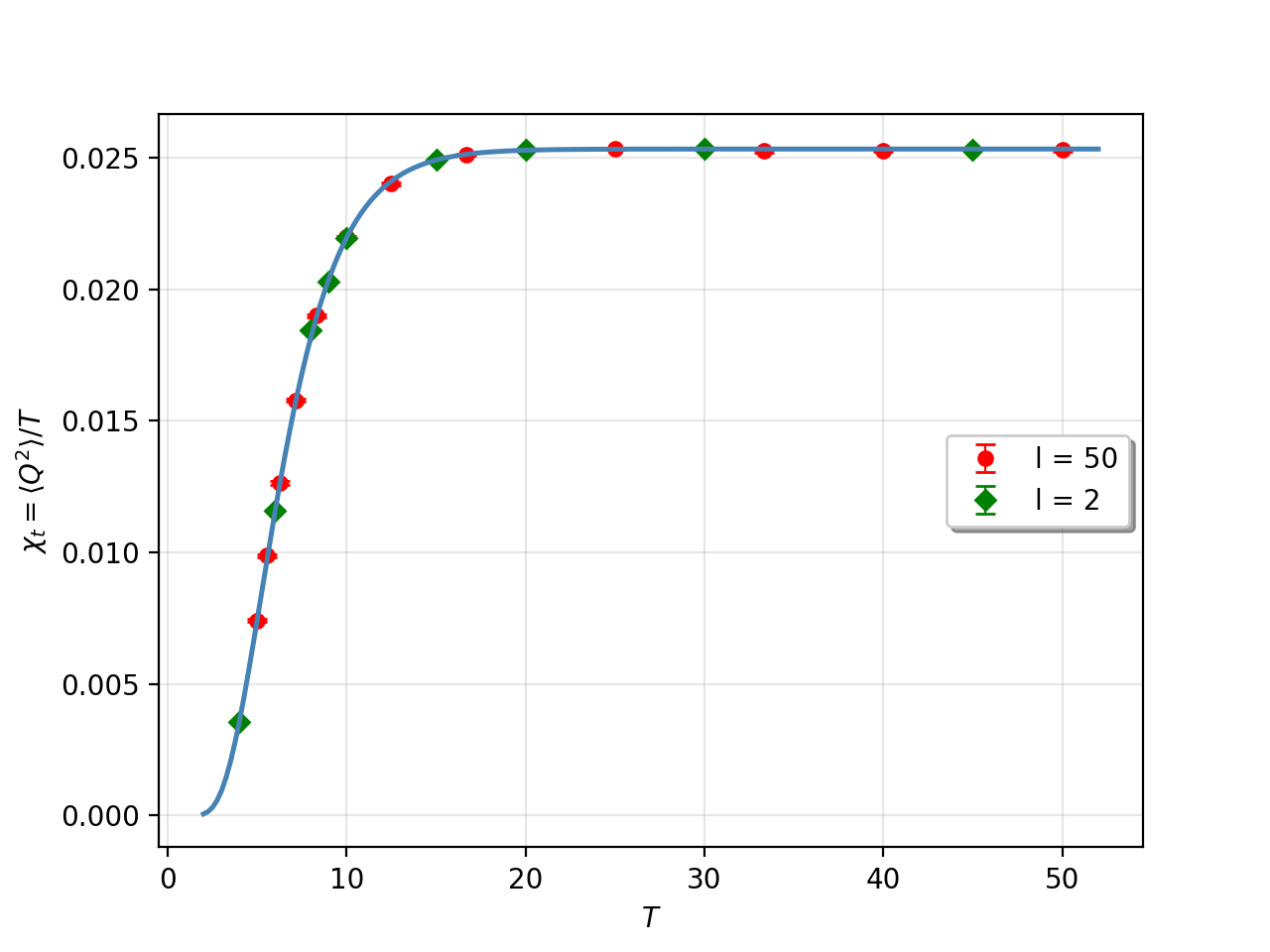}
\caption{Topological susceptibility for different values of the lattice size $T$. The continuous line corresponds to the analytical expression given in \eqref{susceqEL}. One set of data (red) correspond to simulations where all the $l=50$ extended variables were updated in each iteration. The other one (green) where only $l=2$ variables were updated in smaller lattices.}
    \label{susctop}
\end{figure}

The coincidence between the results of the Monte Carlo simulations and the analytical values of the topological susceptibility, for different values of the lattice size $T$, is shown in Fig. \ref{susctop}. Two sets of data are shown: one was obtained using a lattice of $N=50$ nodes, where at each iteration all the variables $x_i$ were changed as described above; the other one corresponds to simulations where at each iteration only two adjacent extended variables were updated. In this case the number of nodes has been kept low, between 2 and 6, and $\delta_t$ adjusted accordingly in order to cover the time range shown; notice that the dependence of the analytic result in \eqref{susceqEL} on $\delta_t$ and $N$ only enters through the combination $T=N \delta_t$. Furthermore, several simulations have been performed with different ranges for the continuous shifts $\delta^\varphi$ of the extended variables. We checked that varying the sizes of the range from which the updates are randomly chosen, between $[-10^{-3}, 10^{-3}]$ and $[-1/2, 1/2]$, the results obtained keep being in a good numerical agreement with the analytical ones.

\begin{figure}
\includegraphics[width=1.0\linewidth]{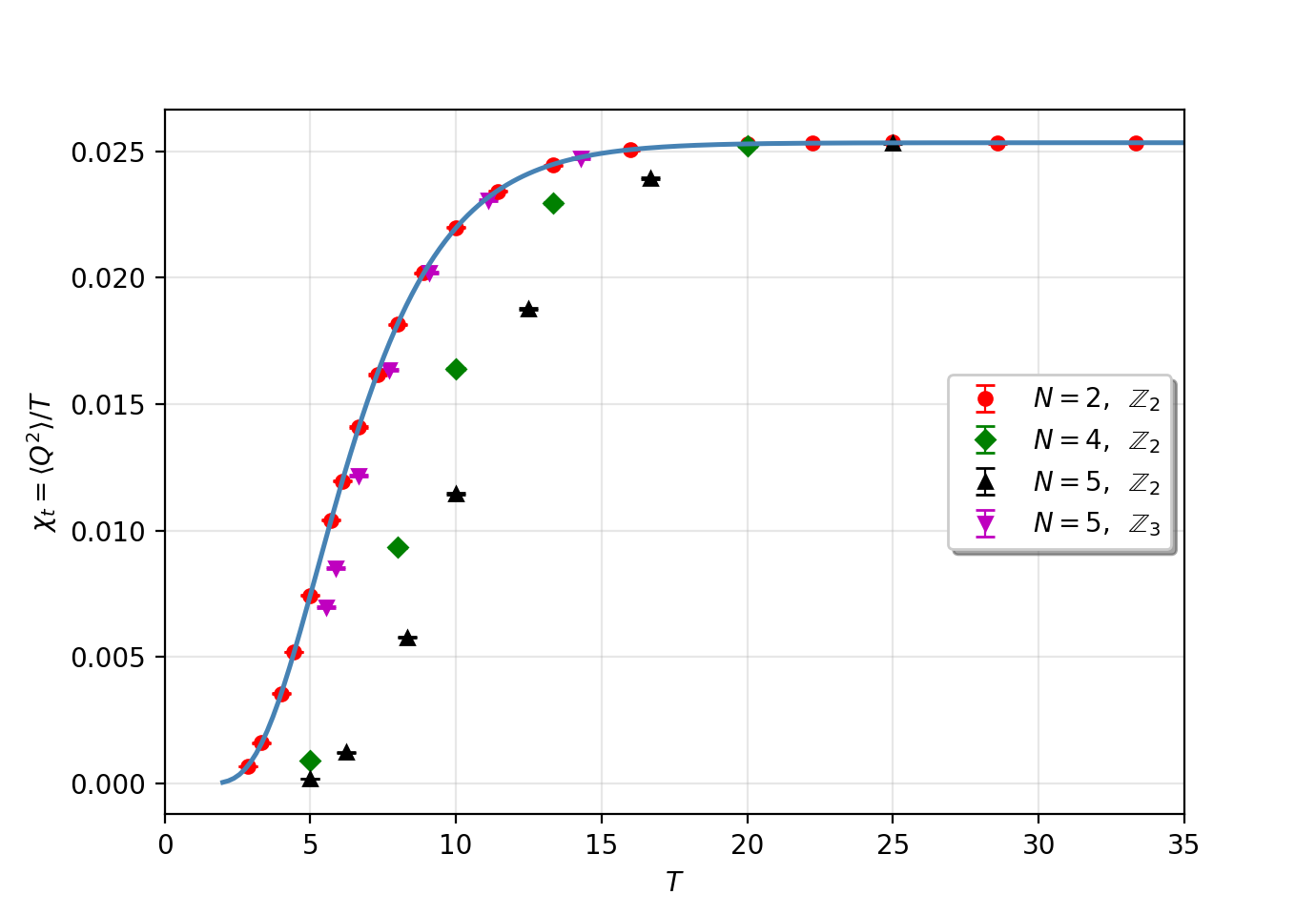}
\caption{Topological susceptibility for different values of the lattice size $T$. The continuous line corresponds to the analytical expression given in \eqref{susceqEL}. The numerical data have been obtained using lattices of different $N$ nodes, where the extended variables take values in $\pi \mathbb{Z}$ ($\mathbb{Z}_2$) and $2 \pi/3 \mathbb{Z}$ ($\mathbb{Z}_3$).}
    \label{suscdiscr}
\end{figure}

In Section \ref{DELFSection} it has been shown that the continuous topological charge distribution is obtained also by restricting the lattice variables to the $U(1)$ discrete subgroups $\mathbb{Z}_n$, as long as $n$ is a multiple integer of the number of lattice nodes $N$. In this case the extended variables $x_i$, which express the relative angular positions, take values in $2 \pi/n \mathbb{Z}$. In the figures presenting our results, this set of variables describing the extended lattice field will be denoted by $2\pi /n \mathbb{Z} (\mathbb{Z}_n)$.  We have checked that this is the case by performing Monte Carlo simulations where the extended variables updates are subject to that restriction, for different values of $n$. In Fig. \ref{suscdiscr} we show some results for the topological susceptibility, when $U(1)$ is approximated by $\mathbb{Z}_2$ and $\mathbb{Z}_3$. The best results correspond to $N=2$ and $\mathbb{Z}_2$, hence satisfying the divisibility condition of $n/N$ being an integer. The remarkable thing of these numerical data is that they are generated by simulations on a lattice with only two nodes, where the extended variables can only take integer multiple values of $\pi$. This makes the algorithm much more efficient and the results still perfectly coincide with the analytical values in the continuum.

\begin{figure}
\includegraphics[width=1.0\linewidth]{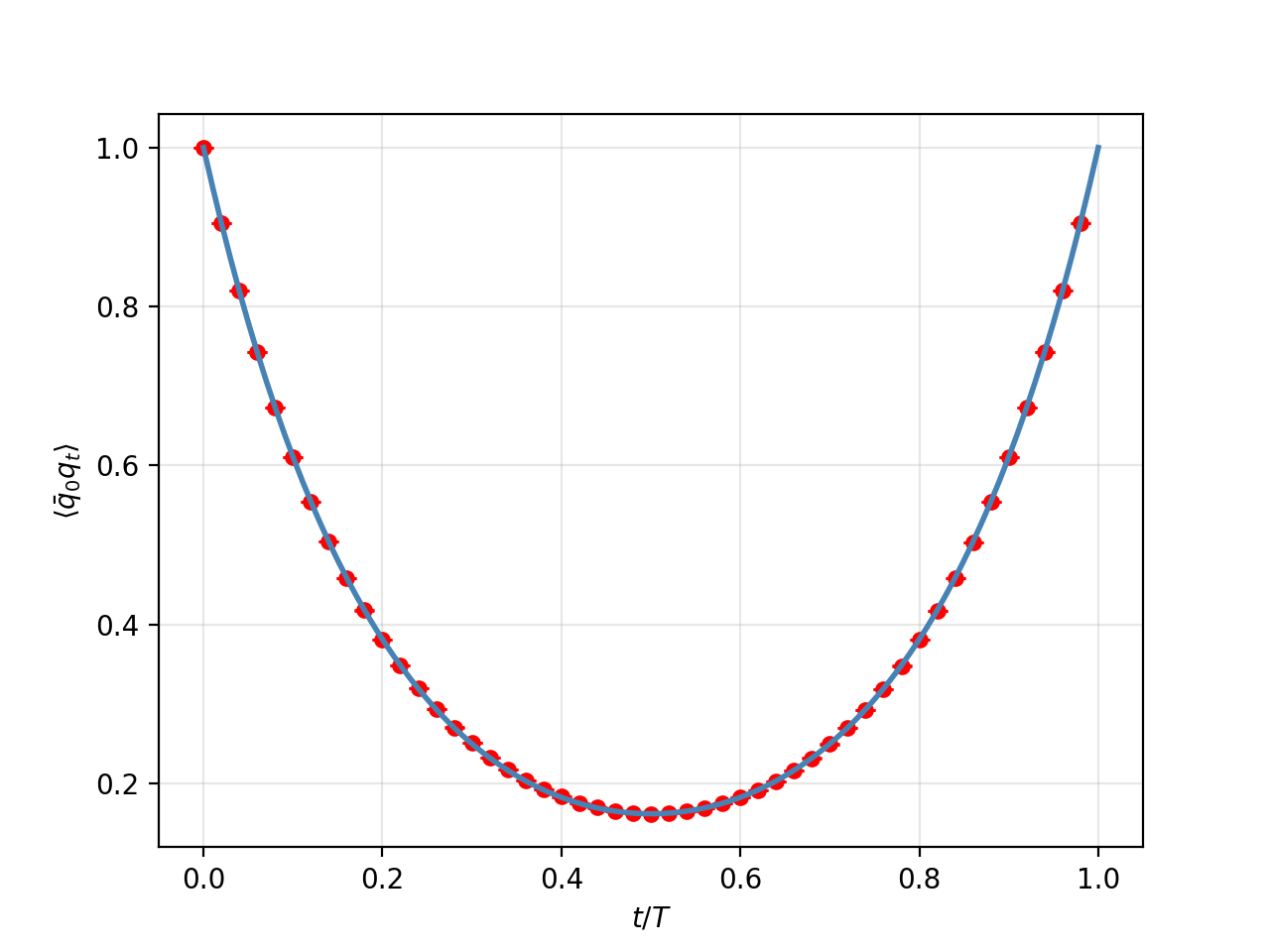}
\caption{Correlation function between two lattice variables as their temporal distance $t$ varies. The Monte Carlo simulations match the analytical expression given in \eqref{correqEL} for $\theta = 0$. The numerical data have been generated on a lattice with $N=50$ nodes and a $\delta_t=0.2$ lattice spacing.}
    \label{corrfig}
\end{figure}

We also numerically measured the values for the correlation function $\langle \bar q_0 q_t \rangle_{\theta, T}$ ($t = l\delta_t$) between the lattice variables, expressed in terms of the extended variables $x_i$: $\langle \bar q_0 q_{l} \rangle_{\theta, T} = \langle e^{i \sum_{i=0}^{l-1} x_i} \rangle_{\theta, T}$. Since this corresponds to the expectation value of an oscillating function, there is a drawback in using large values of $\delta_t$, which came in handy in the simulations of the topological susceptibility: a high frequency of accepted configurations with very different $x_i$, i.e. belonging to different topological sectors, can cause spurious oscillations in the outcome of the expectation value. For the estimation of the correlation function it is therefore better to use fine lattices, as it can be seen in Fig. \ref{corrfig}, where the Monte Carlo data are compared with the analytical expression give in \eqref{correqEL} with $\theta=0$. The simulations have been performed on a lattice with $N=50$ nodes and a lattice spacing of $\delta_t=0.2$. 

Dealing with a quantum perfect action, the size of the lattice spacing should not cause any systematic error in the simulations. We have checked in fact that one can estimate both the topological susceptibility and the correlation function within the same simulation, using for instance a $\delta_t \sim 1$, at the only cost of having to increase the statistics in order to achieve the same agreement with the values in the continuum, going from $\sim 10^7$ iterations to $\sim 10^9$ iterations.

\begin{figure}
\includegraphics[width=1.0\linewidth]{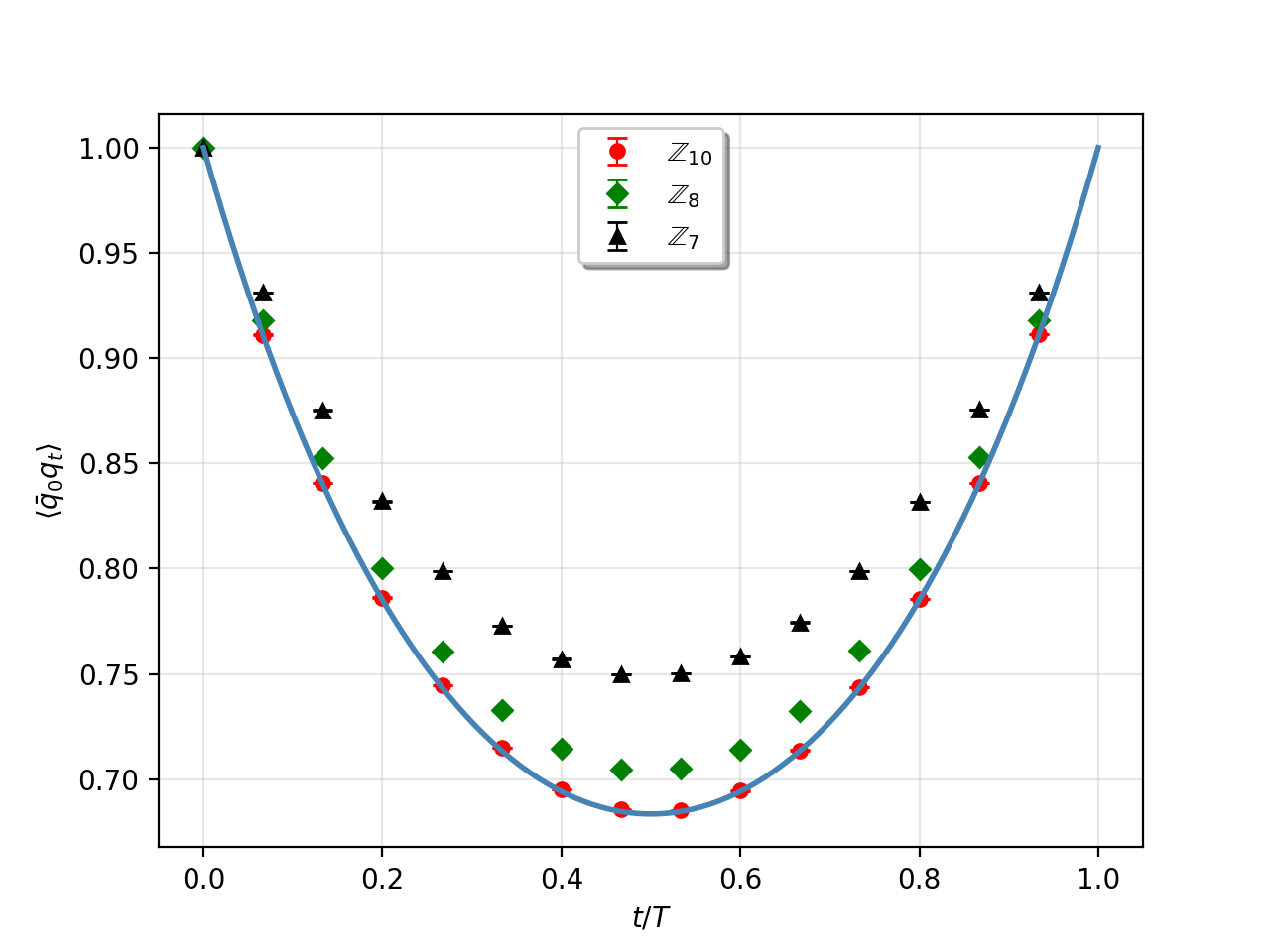}
\caption{Correlation function between two lattice variables as their temporal distance $t$ varies ($\theta=0$). The numerical data have been generated on a lattice with $N=15$ nodes and a $\delta_t=0.2$ lattice spacing. The extended variables are restricted to take values in $2\pi/7 \mathbb{Z}$ ($\mathbb{Z}_{7}$), $\pi/4 \mathbb{Z}$ ($\mathbb{Z}_{8}$) and $\pi/5 \mathbb{Z}$ ($\mathbb{Z}_{10}$) .}
    \label{corr_discr}
\end{figure}

The correlation function can also be estimated using the discrete subgroups $\mathbb{Z}_n$ as domain of the lattice variables, as done with the topological susceptibility. In this case, however, there is some loss of information due to the identification of different sets energy eigenstates that this restriction implies, as explained in section \ref{DELFSection}. 
In this example, a good numerical precision is achieved for sufficiently large values of $n$. In Fig. \ref{corr_discr} data are shown from simulations on a lattice with $N=15$ nodes, where the lattice variables take values in different $\mathbb{Z}_n$. As it can be seen, the numerical values approach the analytic curve in the continuum as $n$ grows. 
However, it is important to notice that Lie groups like $SU(n)$ have only a finite set of discrete subgroups. Thus, a limit in which the subgroup becomes finer is possible for $U(1)$, but it is not possible in other important examples. 

The proposal to use extended lattice fields restricted to discrete subgroups introduced in Section \ref{DELFSection} is potentially useful to study topological aspects of the field only. 
For example, we ran a simulation with $10^7$ iterations using $\Z_8 \subset U(1)$ and $(N=4, \delta_t = 1)$. For the correlation function with time difference of $2$ time steps, the deviation from the analytical value for the continuum was of $4\%$. On the other hand, the numerical measurement of the topological susceptibility agreed with the analytical result in the continuum up to $0.03\%$.

\section{General description of extended lattice fields and complementary examples}\label{GeneralEFSection}

In dimension $d$, an ${\cal F}$-valued extended lattice field $\phi^{EL} \in {\cal M}_{EL}$ on $(M, C)$ a base space endowed with a cellular decomposition (which may be a cubiculation, or a triangulation) consists of $d+1$ compatible maps $\phi^{EL}= ( \phi^0, \ldots , \phi^d )$. The map $\phi^0 = \phi^L: C^0 \to {\cal F}$ is the standard lattice field, and the map $\phi^{i+1} : C^{i+1} \to \mbox{Ext}(\phi^0, \ldots , \phi^i)$ stores homotopic information about the restriction of the field to each of the $(i+1)$-dimensional cells in $C$ relative to the evaluation of the maps $( \phi^0, \ldots , \phi^i )$.%
\footnote{
This tower of compatible maps, where the first one is the standard lattice field can be interpreted as a collection of fields. The new extra fields have a given topological interpretation which implies that their possible joint evaluations are not free but subject to constraints, as we will show in Subsection \ref{2dSs}. 
}
 
It is clear that ${\cal M}_{EL}$ is a covering space of the space of standard lattice fields ${\cal M}_L$. Every instance of a nontrivial map $\phi^i$  indicates a nontrivial fiber of the covering map, and an action of $\pi_i {\cal F}$ (or several copies of this group) on the fiber which changes the topological sector of the extended lattice field. 

If we start with a field in the continuum $\phi^{cont}$ and evaluate the maps $( \phi^0, \ldots , \phi^d )$ on all the cells of $C$, we have the corresponding extended lattice field $\phi^{EL}$, which has complete knowledge of the homotopy class of $\phi^{cont}$. Thus, any topological charge that in the continuum may be evaluated as a winding number is computable in terms of extended lattice fields, and the space of extended lattice fields is decomposed in a set of connected components which is in 1 to 1 correspondence with the set of connected components of the space of fields in the continuum. 
This qualitative result and the whole framework of extended lattice fields for sigma models have a counterpart for gauge fields \cite{MZ1, MZ2}.

Notice that an extended lattice field is not just a standard lattice field together with one integer number representing ``its topological charge''; all the information contained in the field is of local character. In topology the homotopy groups $\pi_k X$ contain information about global aspects of the space $X$, and the definition of the spaces $\pi_k X$ is global in essence. On the other hand, topology is based on local information. The question of how to determine $\pi_k X$ for a space $X$ which is decomposed into pieces, like a cellular decomposition, in terms of information of the individual pieces and gluing information, was answered long time ago by Seifert and van Kampen for the case of the fundamental group $\pi_1 X$ (which is everything we care about in our simple example). More recently, Brown and Higgins have formulated a higher homotopy Seifert van Kampen theorem \cite{BH78, Seifert31, Kampen33} (which is the relevant result for general extended lattice fields).

A discrete extended lattice field restricts the possible evaluations of the map $\phi^0$ to a discrete subgroup $H \subset G$, and keeps all the homotopical information.

\subsection{Another 1-dimensional example: the rigid body}

In this section we study the topological properties of another 1-dimensional model, the quantum rigid body, whose configurations at each time are described by elements of $SO(3)$. This system is a nonlinear nonabelian sigma model, but even when it is nontrivial from this perspective, there are other points of view from which it is a simple system: At the classical level it is known to be an integrable system whose integrability is preserved by a wise time discretization \cite{Veselov88}, and at the quantum level the heat kernel action is a quantum perfect action for it. 

Ordinary lattice fields are assignments of $SO(3)$ elements to the vertices of our time lattice. Extended fields contain this information, and they also tell us the homotopy class of a curve in $SO(3)$ associated to each elementary time subinterval $[t_i, t_{i+1}]$. 
As we saw earlier, this information may be conveniently stored in a ``relative difference'' variable valued in the universal cover of the group, which for $SO(3)$ can be identified with $SU(2)$: 
Given a time interval $[t_i, t_{i+1}]$ and an evaluation of the lattice field $q_i , q_{i+1} \in SO(3)$ at its end points, the ``relative difference'' variable $x_{i, i+1}= x_i$ satisfies $q_{i+1} = q_i \pi(x_i)$. Notice that once the boundary values $q_i , q_{i+1} \in SO(3)$ are given, there are two elements of $SU(2)$ associated to $[t_i, t_{i+1}]$ which satisfy the condition. Each of them corresponds to a different homotopy class of histories in the continuum. 
The extended field content restricted to a link can thus be described by 
\begin{equation}
\left. \mathcal{M}_{EL} \right |_{[t_i, t_{i+1}]} \simeq SO(3) \times SU(2).
\end{equation}
When the system is subject to periodic boundary conditions, histories separate in two classes corresponding to the fundamental group of our configuration space $\pi_1(SO(3)) \simeq \mathbb{Z}_2$: Contractible histories are associated with a topological charge $Q=0$, and non-contractible histories are associated $Q=1$. For a lattice with $N$ links, the total extended lattice space is 
\begin{equation}
\mathcal{M}_{EL} = \mathcal{M}_{EL}(Q=0) \sqcup \mathcal{M}_{EL}(Q=1),
\end{equation}   
where
\begin{equation}
\mathcal{M}_{EL}(Q) \simeq SO(3)\times SU(2)^{N-1}.
\end{equation}
The $x_i \in SU(2)$ extended variables associated to the trivial $Q=0$ sector, i.e. to the contractable histories, satisfy $\prod_{i=1}^N x_i = \mathbb{1}$; while the variables associated to the $Q=1$ sector satisfy $\prod_{i=1}^N x_i = \mathbb{-1}$ and describe non contractible trajectories in $SO(3)$. 

We will use the heat kernel measure to model the dynamics of the system. The Boltzmann factor of a history is factorized as contributions associated to the time subintervals, and for each interval it is given by 
\begin{equation}\label{Boltzmann}
e^{- S_{HK}(x_i)} = \sum_{n=0}^{\infty} d_n \chi_n(x_i)e^{-C_n \delta_t} =  \sum_{n=0}^{\infty} (n+1)\frac{\sin((n+1)\theta_i)}{\sin(\theta_i)}e^{-\frac{n(n+2) }{4} \delta_t},
\end{equation}    
where the integers $n$ label the irreducible representations of $SU(2)$, $d_n = n+1$ is the dimension of the representation, $C_n = n(n+2)/4$ is the Casimir operator and $\chi_n(x_i)$ is the character of $x_i$ in the n-representation. In the last equality we have used  Weyl's formula for the character, in which $\theta_i = \arccos\left(\chi_1(x_i)/2 \right)$, with $\chi_1(x_i)$ the trace of the element $x_i$ in the fundamental representation.

The expectation value of the topological charge is therefore given by
\begin{equation}
\langle Q \rangle = \frac{1}{Z}\int \prod_{l=1}^N dx_l \,\delta_{\prod_{l} x_l, \mathbb{-1}} \prod_{l=1}^N \sum_{n=0}^{\infty} (n+1) \chi_n(x_l) e^{-\frac{n(n+2)}{4}\delta_t},
\end{equation}
where the partition function $Z$ is
\begin{equation}
\begin{split}
Z &= \int \prod_{l=1}^N dx_l\, \delta_{\prod_{l} x_l, \mathbb{1}} \prod_{l=1}^N \sum_{n=0}^{\infty} (n+1) \chi_n(x_l) e^{-\frac{n(n+2)}{4}\delta_t} \ +\\
&\int \prod_{l=1}^N dx_l\, \delta_{\prod_{l} x_l , \mathbb{-1}} \prod_{l=1}^N \sum_{n=0}^{\infty} (n+1) \chi_n(x_l) e^{-\frac{n(n+2)}{4}\delta_t}.
\end{split}
\end{equation}
The integrals can be performed by expressing the delta functions in terms of their character expansion, $\delta_{\prod_{l} x_l, \pm \mathbb{1}} = \sum_m \beta^{(\pm)}_m \bar \chi_m\left (\prod_{l} x_l \right)$, with the coefficients $\beta^{(\pm)}_m = (\mp 1)^m (m+1)$, and using the orthogonality of the characters. The final result is 
\begin{equation}
\label{charge}
\langle Q \rangle = \frac{\sum_{n = 0}^\infty (-1)^n (n+1)^2 e^{- \frac{n(n+2) T}{4}}}{ \sum_{n=0}^\infty (1+(-1)^n)(n+1)^2 e^{- \frac{n(n+2) T}{4}}},
\end{equation}
where $T= N \delta_t$ is the temporal length. 

\begin{figure}
\includegraphics[width=1.0\linewidth]{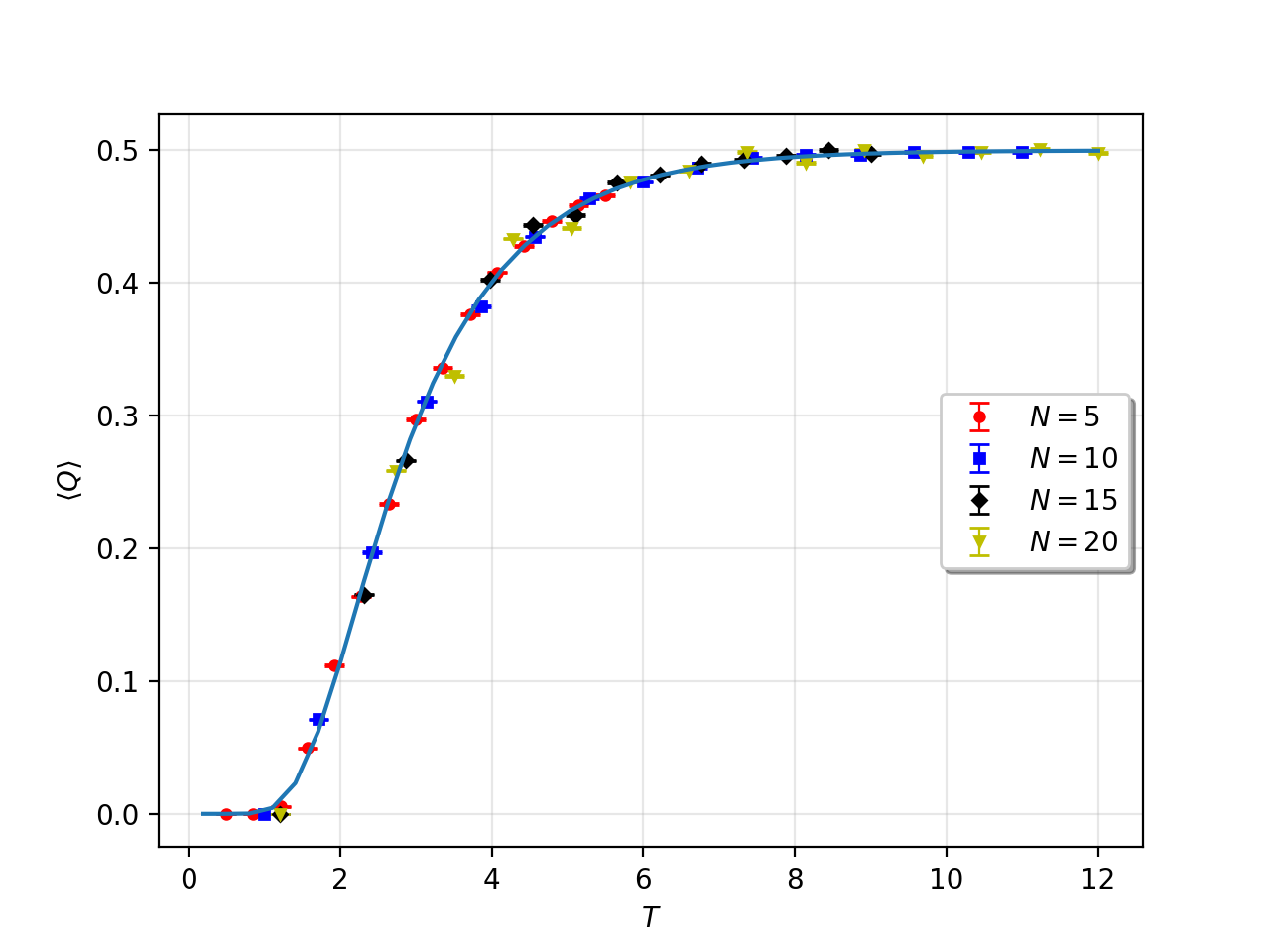}
\caption{Expectation value of the topological charge for the quantum rigid body. The continuous line corresponds to eq. \eqref{charge}. Simulations are shown for lattices with different number $N$ of vertices.}
    \label{chargefig}
\end{figure}

We also evaluated $\langle Q \rangle$ via Monte Carlo simulations, using again the traditional Metropolis algorithm. Notice that equation \eqref{Boltzmann} for the Boltzmann factor associated to a time subinterval depends only on the variables $\theta_i$ and $\delta_t$. In order to speed up the simulations, we stored the values of $e^{-S_{HK}}$ corresponding to a 2-dimensional partition in the variables $\theta  \in [0,\pi]$ and $\delta_t \in [0.05, 2]$ and used a cubic interpolation of these values when applying the Metropolis algorithm. The results are shown in Fig. \ref{chargefig}, where the data refer to lattices with different number $N$ of vertices. The fact that these data all coincide at any fixed value of $T$, i.e. for different values of $\delta_t$, corroborates that the lattice model is quantum perfect. 

The algorithm is similar to the one described in the previous section. In this case, however, the potential new configurations are generated by uniformly sampling from the compact space of $SU(2)$ matrices. Since this procedure may cause large changes in the heat-kernel weight function, only two adjacent extended variables are updated at each iteration. We start from an initial configuration where all the elements coincide with the identity matrix. 
At each iteration it is decided if the proposed update will change the topological charge or not. Then an element $x_i$ is selected and a potential new element $x'_i = x_i \, r$ is generated, $r$ being a random $SU(2)$ matrix. A potential new contiguous element $x'_{i+1}$ is also defined. In the case that the proposal preserves the topological charge we define $x'_{i+1} = r^{-1}\,x_{i+1}$. Notice that  $x'_{i}\, x'_{i+1} =x_{i}\, x_{i+1}$. When the proposed new configuration shifts the value of the topological charge, we define $x'_{i+1} = - r^{-1}\,x_{i+1}$, leading to $x'_{i}\, x'_{i+1} = - x_{i}\, x_{i+1}$. The updated configuration is then passed to the Metropolis acceptance test.

\begin{figure}
\includegraphics[width=1.0\linewidth]{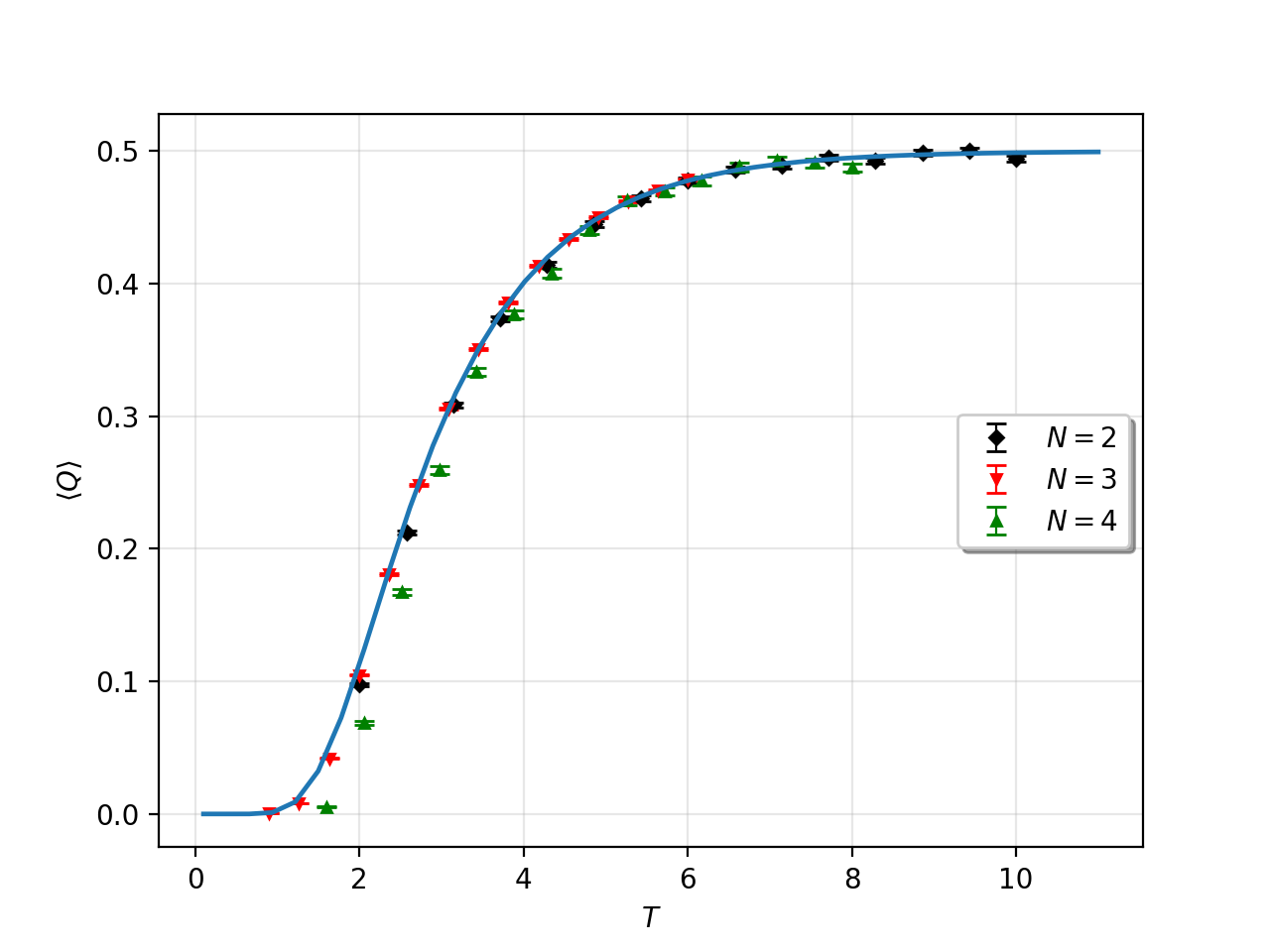}
\caption{Expectation value of the topological charge for the quantum rigid body, when the extended variables are restricted to the binary tetrahedral subgroup, on lattices with $N=2, N=3$ and $N=4$ vertices. The continuous line corresponds to eq. \eqref{charge}. }
    \label{tetrafig}
\end{figure}

We also used discrete extended lattice fields to study the expectation value of the topological charge. The discrete group used was ``the binary tetrahedral group'' $2T \subset SU(2)$ which is the double cover of the tetrahedral group $T \subset SO(3)$. It is constituted by 24 elements:
\begin{equation}
\left \{\pm \mathbb{1},  \pm i\sigma_x, \pm i\sigma_y, \pm i \sigma_z, \frac{1}{2}\left(\pm \mathbb{1} \pm i \sigma_x \pm i \sigma_y \pm i \sigma_z \right) \right \},
\end{equation}
where $\sigma_x, \sigma_y, \sigma_z$ are the Pauli matrices. The results of the simulations are shown in Fig. \ref{tetrafig}.

Our experience with the $\Z_n \subset U(1)$ sigma model tells us that, when studying the topological susceptibility, we should expect to find agreement between the $T \subset SO(3)$ and the $SO(3)$ sigma models only under special circumstances. We do not yet know a set of conditions telling us when we should expect agreement. In Figure \ref{tetrafig} we present results showing that in several simulations we have complete agreement (up to the accuracy of our simulations) and in other simulations there is some deviation.

\

\subsection{Extended lattice fields in two dimensions}\label{2dSs}

We already presented two 1-dimensional examples. Now we describe the space of extended fields for two 2-dimensional examples. In the first example, the base space will be $(M = T^2 , C)$ the torus with a cartesian grid $C$ subdividing it into squares (or plaquettes), and the field will be valued in ${\cal F} = U(1)$. Let us focus in the restriction of the field to a given plaquette $P$ with vertices $\{ v_i \}$ with $i \in \{ 1, \ldots , 4 \}$ and links $\{ l_{ij} \}$ labeled by pairs of neighboring vertices. 
The general definition of extended fields tells us that 
${\cal M}_{EL}|_P \subset U(1)^4 \times \mbox{Ext}(l_{12}) \times \mbox{Ext}(l_{23}) \times \mbox{Ext}(l_{34}) \times \mbox{Ext}(l_{41})$, where the description of the subset must guarantee that there are fields in the continuum restricted to $P$ with the prescribed extensions $\mbox{Ext}(l_{ij})$. 
Our experience with the 1-dimensional case presented in Section \ref{extended} tells us that we can describe the 1-dimensional contribution of the extended field using a map $x : C^1 \to \widetilde{U(1)} = \R$ complementing $\phi^0$. Recall that $x(l_{ij}) = - x(l_{ji})$, and that in order to simplify notation we write $x_{ij}= x(l_{ij})$. Then 
$\phi^0(v_j)= \phi^0_j = e^{i x_{ij}} \phi^0_i$. 
There is an important difference, however, we need to guarantee that the data given at $\partial P$, the boundary the plaquette, is compatible with the existence of a field in the interior of $P$. This requirement implies that 
\begin{equation}\label{XclosureAtP}
	\sum_{(ij)< \partial P} x_{ij} = 0 . 
\end{equation}
The map $q^2$ needed to complete the description of the extended field is trivial in this case because there is only one homotopy class of maps from the square to the circle relative to fixed boundary conditions. 
Thus, an abbreviated description of the degrees of freedom of the field restricted to $P$ may be given as $\phi^{EL}|_P = (\phi^0(v_1); \{ x_{ij} \})$, where condition (\ref{XclosureAtP}) is required to hold. Then we can write ${\cal M}_{EL}|_P \simeq U(1) \times \R^3$. 

Let us now consider the whole space instead of a single plaquette. 
Given a noncontractible loop $l$, we require that $\sum_{(ij)<l} x(l_{ij}) = 2\pi n$ for some $n \in \Z$; this integer is a topological charge. 
We can select two noncontractible loops $l(1), l(2)$
generating the fundamental group of $T^2$, in order to write two independent topological charges $Q_I(\phi^{EL}) = \sum_{(ij)<l_I} x(l_{ij}) = 2\pi n_I$. 
The space of extended fields with given topological charges is 
\begin{equation}
	{\cal M}_{EL}(n_1, n_2) \simeq U(1) \times \R^{N_1 - (N_2 - 1) - 2} , 
\end{equation}
where $N_2-1$ indicates that among the $N_2$ constraints (\ref{XclosureAtP}) there is one redundancy, and the subtraction of $2$ appears because we have specified the value of two topological charges. 
Accordingly, the space of extended fields in this example is 
\begin{equation}	
{\cal M}_{EL} = \sqcup_{n_1, n_2} {\cal M}_{EL}(n_1, n_2) . 
\end{equation}
This is an explicit presentation of the space of extended fields as a disjoint union of topological sectors. 

We can use our experience with the 1-dimensional example presented in Section \ref{extended} to give expressions for two topological charges. It is also simple to write an action regularizing the free field valued in $U(1)$. The resulting model will not have a quantum perfect action, and its study would be delicate. 

In the Appendix, we present a description of a 1-dimensional extended lattice $U(1)$ field using a pair $(q^L, z^{(u)})_u$ of a standard lattice $U(1)$ field and a standard $\Z$-lattice gauge field, but it all has to be done within local trivializations of the space of extended fields labeled by $u \in U(1)$. It is clear that the same considerations can be used in the 2-dimensional example considered now,%
\footnote{
If our map $\phi^2$ contained nontrivial information, we would need three fields ab initio. 
} 
the only extra element is equation \eqref{XclosureAtP}. When translated into the mentioned variables, and within a local trivialization, equation \eqref{XclosureAtP} translates into requiring that the $\Z$-lattice gauge field be flat. 
Interestingly, if we use the picture where $z^{(u)}$ is a field in the dual lattice, describing $\Z$-decorated oriented curves in the dual lattice, equation \eqref{XclosureAtP} becomes a sort of conservation law for the $\Z$-decorated oriented curves. The methodology used in references \cite{AGGST19, NTU21, SG19, GGS18, SGG20, Sulej20} implements the flatness constraint originated by equation \eqref{XclosureAtP} by introducing yet another field appearing as a ``Lagrange multiplier''%
\footnote{
In higher dimensions, the mentioned redundancy in the system of equations resulting from imposing \eqref{XclosureAtP} for each plaquette would lead to a another field associated to 3-dimensional cells. 
}

In Section \ref{DELFSection} we developed a framework for extended lattice fields in which the part corresponding to the standard lattice field was restricted to a discrete subgroup, we called those fields discrete extended lattice fields. 
The same 2d system described above 
could be described using discrete extended lattice fields with $\phi^0$ valued in $\Z_n \subset U(1)$. 
All the remarks about the space of fields and the calculation of the action and topological charges simply by restriction of the formulas for extended fields to the fields with $\phi^0$ valued in $\Z_n$ given in Section \ref{DELFSection} apply to this example.

In the second 2-dimensional example we only describe the fields and the space of fields at the local level. The base space $(M = T^2 , C)$ is the torus (with a cartesian cellular decomposition), and the target space ${\cal F} = S^2$ is the sphere. 
Again, an extended field consists of a triple of maps $q^{EL}= ( q^0, q^1 , q^2 )$, but in this case $q^1$ contains no information (because in $S^2$ any two curves with the same end points are homotopic), and $q^2$ is nontrivial. Moreover, the evaluation of the map $q^2$ on the different $2$-cells is not constrained. 
Then, when restricted to a plaquette $P$, the space of extended fields is ${\cal M}_{EL}|_P = (S^2)^4 \times \mbox{Ext}(P) \simeq (S^2)^4 \times \Z$. 
We know that ${\cal M}_{EL}$ is a covering space for ${\cal M}_L = (S^2)^{N_0}$, and that ${\cal M}_{EL}$ is a disjoint union of topological sectors indexed by their topological charge labeled by the integer numbers (because $\pi_2(S^2) = \Z$).

\section{Summary, qualitative comparison and outlook}
\label{Summary}

We introduced an extended notion of lattice field for sigma models which includes local homotopy data as part of the field, in analogy of the treatment of lattice gauge fields in references \cite{MZ1, MZ2}. 
Even when the target space is connected, the resulting space of fields is not in general a connected space in contrast with the space of standard lattice fields. Extended lattice fields are well suited to model topological aspects of the field without making an excision in the space of fields, as is needed in the case os standard lattice fields. In the introduction, we mentioned a list of shortcomings of standard lattice fields with respect to the study of topological aspects of the field. Below we review how our proposal performs with respect to two of those issues: 

We showed that there is a 1 to 1 correspondence between the set of topological sectors of the space of fields in the continuum and that of the space of extended lattice fields. Thus, extended lattice fields model all topological sectors even if the model is not near the continuum limit. 

We also showed models, capable of modeling topological aspects of the field, which have a positive definite transfer matrix. This is in contrast with generic behavior of models using standard lattice fields after the excision needed to model topological aspects \cite{GK82, Creutz04}. 

In Section \ref{extended} we showed that the heat kernel measure was appropriate for extended lattice fields, and we also saw that integrating over the extra (topological) degrees of freedom yielded the heat kernel measure (or Villain measure) for standard lattice fields. This was shown for the simple 1-dimensional $U(1)$ model, but it is a general property. A proof that this property holds in general follows from the fact that the space of extended lattice fields is a covering space of the space of standard lattice fields. 
We argued that working with the Villain measure and letting the ``integer valued Villain variables'' play a role with respect to topological aspects of the field is equivalent to using extended lattice fields. Interestingly, from this point of view extended lattice fields are not new for abelian theories. One instance in which they played a prominent role in gauge theories was the mathematical proof that compact QED without fermions has a non confining fase \cite{BMK77, Guth80, PPW91}. Additionally, it was recently proposed to use  ``Villain's integer valued variables'' to further study topological aspects of abelian fields and the coupling of fermions in the lattice (see for example  \cite{AGGST19, NTU21, SG19, GGS18, SGG20, Sulej20}). This paragraph intends to convince the reader that extended lattice fields, in the context of abelian fields, could be considered as something familiar. The new element is the general framework, which applies also to nonabelian fields and to base spaces of arbitrary dimension.  

Extended lattice fields let us implement a traditional Metropolis Monte Carlo algorithm in simulations that efficiently sample the space of fields without getting stuck in a single topological sector. The results of this first study, presented in Sections \ref{SimulationSection} and \ref{GeneralEFSection}, merely show that it is a possible methodology to study topological aspects of the field. Clearly, it can be refined in several ways; this is just a first study.  


We also introduced a version of extended lattice fields in which all the topological information of extended fields is retained, but where the evaluation of extended fields is restricted to a discrete subset of the target space while the symmetry group is restricted to a discrete subgroup accordingly. 
The proposal is to consider the discrete version of extended lattice fields to study topological aspects of the field only. 
Interestingly, in the abelian 1-dimensional model, which is not a quantum perfect model, the exact topological susceptibility could be evaluated if certain divisibility conditions were met. We showed that the same issue played an important role for the nonabelian model presented in Section \ref{GeneralEFSection}. In that case we did not give precise conditions guaranteeing that the evaluation of the topological susceptibility of the discrete extended model agrees with the continuum; we just presented a first qualitative study. More work is needed to be able to evaluate if this new idea is promising in higher dimensional non abelian systems; the attractiveness of the proposal is that using discrete subgroups could yield very efficient simulations.

\section*{Appendix}

{\em The 1-dimensional $U(1)$ extended lattice field and its relation with a pair composed of a standard $U(1)$-lattice field and a $\Z$-lattice gauge field:}

Consider $\tilde{q}^{EL}, q^{EL}$ two extended lattice fields on the same fiber of the covering map $Pr^{-1}(q^L)$; in other words, those two extended fields induce the same standard field $Pr(\tilde{q}^{EL}) = Pr(q^{EL}) = q^L$. ``The difference'' of these two fields associates an integer to every link $[t_i, t_j] \in L^1$, which can be calculated as 
$q^1_{ij} \circ (\tilde{q}^1_{ij})^{-1} \in \pi_1 (U(1)) = \Z$. 
Let us explain the previous expression: 
$q^1_{ij}, \tilde{q}^1_{ij} \in \mbox{Ext}(q_i , q_j)$ are homotopy classes of curves in $U(1)$ relative to having $q_i$ and $q_j$ as endpoints. We can chose $q|_{[t_i, t_j]}, \tilde{q}|_{[t_i, t_j]}$ a representative of each class, and calculate $[ q|_{[t_i, t_j]} \circ (\tilde{q}|_{[t_i, t_j]})^{-1}] \in \pi_1 (U(1))$. It can be verified that the calculated homotopy class of a based closed curve is independent of the chosen representatives; this is the meaning of our previous expression. 

We could aim to specify an extended lattice field by a pair consisting of a standard lattice field and a ``difference field'' $z: L^1 \to \Z$ (which would be a more concrete realization of our map $q^1: L^1 \to Ext(q^0)$); the only missing ingredient is a ``reference extended field'' $q_0^{EL}(q^L)$ for each standard field $q^L$. 
In other words, we would like to prescribe a parametrization of our spaces $\mbox{Ext}(q_i , q_j)$ by $\Z$. 
This amounts to choosing a trivialization of ${\cal M}_{EL}$ as a covering space of ${\cal M}_L$ . 
Unfortunately, there is no global trivialization or global section $q_0^{EL}(q^L)$ serving as a global ``reference extended field''. 
What we can do is to have a collection of local reference fields and associated local trivializations of ${\cal M}_{EL}$. 
Given any $u \in U(1)$, consider the open proper subset ${\cal M}^{ij}_L(u) = \{ q^L \in {\cal M}_L|_{[t_i, t_j]} : q_i \neq u , q_j \neq u \}$. Notice that ${\cal M}_L|_{[t_i, t_j]}$ is covered by the collection of open subsets ${\cal M}^{ij}_L(-1)$, ${\cal M}^{ij}_L(e^{i 2\pi / 6})$, ${\cal M}^{ij}_L(e^{-i 2\pi / 6})$. In each of these open subsets ${\cal M}^{ij}_L(u)$ we can construct a corresponding section $q^{EL}_{(u)}=(q^0, q^1_{(u)})$, and an induced local trivialization for $Pr^{-1} ({\cal M}^{ij}_L(u))$ which will use coordinates $(q_i \in U(1)-\{u\}, q_j \in U(1)-\{u\}; z_{ij}^{(u)} \in \Z)_u$.%
\footnote{ 
Our choice of relative homotopy class $q^1_{(u)}$ for $q^{EL}_{(u)}|_{[t_i, t_j]} = (q_i, q_j, q^1_{(u)})$ requires that the curve starting at $q_i$ and finishing at $q_j$ has linking number equal to zero with $u$, but choosing other linking number $n \in \Z$ is equally valid. 
} 
For two different local trivializations the evaluation of the ``difference field'' may differ; for example, given $q^{EL} \in Pr^{-1} ({\cal M}^{ij}_L(-1)) \cap Pr^{-1} ({\cal M}^{ij}_L(e^{i 2\pi / 6}))$ we can calculate that 
$z_{ij}^{(e^{i 2\pi / 6})} = z_{ij}^{(-1)} + q^1_{(-1)} \circ (q^1_{(e^{i 2\pi / 6})})^{-1}$. 
The following formula relates this description to the variables $(q_i \in U(1), x_{ij} \in \R)$ used in the rest of the article 
\begin{equation}\label{qq}
	q^{EL}|_{[t_i, t_j]}= 
	(q_i, x_{ij}) \mapsto (q_i, q_j = q_i e^{i x_{ij}}; 
	z_{ij}^{(u)} = \frac{1}{2\pi}(x_{ij} - x_{ij}^{(u)} )_u) , 
\end{equation}
where we have used the reference section written as $q^{EL}_{(u)}|_{[t_i, t_j]} = (q_i, x_{ij}^{(u)})$. 
At each local trivialization, this formula has an inverse 
\begin{equation}
	(q_i, q_j ; z_{ij}^{(u)} )_u \mapsto 
	(q_i, x_{ij}= \delta (u, q_j) - \delta (u, q_i) + z_{ij}^{(u)} 2 \pi), 
\end{equation}
where $\delta (u, q_j = u e^{it}) = t$ for all $t \in (0, 2\pi)$. 

As far as the topological charge is concerned, it is interesting to notice that if the evaluation of the field $q^{EL}$ at every vertex of the time lattice is different from $u \in U(1)$, then we can describe the field using our choice of $u$-local trivialization $q^{EL}= (q^L, z^{(u)})_u$, and calculate the topological charge as 
\begin{equation}\label{Q(z)}
	Q_{EL}(q^L, z^{(u)})_u = \sum z_{ij}^{(u)} . 
\end{equation}

Notice that within the local trivialization associated to $u$, the new field $z^{(u)}$ behaves like a lattice $\Z$-gauge field. 
Let us consider a pair of lattice links $(ij), (jk)$. Our extended field lets us distinguish between different relative homotopy classes of fields in the time interval $[t_i, t_k]$ which lead to the same lattice field $q^0|_{[t_i, t_k]} = (q_i, q_j, q_k)$; the extra data stored in the extended field consists of two integers $q^1|_{[t_i, t_k]} = (z_{ij}^{(u)}, z_{jk}^{(u)})$. Now recall that these two integers are relative to an arbitrary choice of reference section, and other choices are equally valid. It is possible to change the reference sections in both lattice links in a way that the compounded variable $z_{ij}^{(u)} + z_{jk}^{(u)}$ does not change; the way to do it is to set $(z_{ij}^{(u)})' = z_{ij}^{(u)} + n_j^{(u)}$ and $(z_{jk}^{(u)})' = z_{jk}^{(u)} - n_j^{(u)}$. 
This corresponds to a gauge transformation at vertex $j$. 
The action of this gauge transformation can be thought of as changing the convention defining $q_{(u)}^{EL}(q^L)$ as having linking number equal to zero with $u \in U(1)$; the new convention would be that the linking number of $q_{(u)}^1([t_i, t_j])$ with $u$ be $n_j^{(u)}$ and that the linking number of $q_{(u)}^1([t_j, t_k])$ with $u$ be $-n_j^{(u)}$. 
(An alternative image is that, within each local trivialization, $z^{(u)}$ is a standard lattice $\Z$-valued field living on the dual lattice.) 

To finalize this comment, we have to mention the measure in ${\cal M}_{EL}$ described in terms of these variables. We had already mentioned that the measure in $\mbox{Ext}(q_i , q_j)$ was the counting measure. This is compatible with our description in which $z^{(u)}$ is a $\Z$-gauge field. Once the reference fields are chosen, the gauge field is $\Z$-valued, and the measure is the counting measure. We also used this measure in a calculation; see equation \eqref{T_ELdiagonalized} and the previous discussion. 

We wonder if this concrete realization of the extended lattice field $q^{EL}= (q^0, q^1)$ as a pair of fields where $q^0 = q^L$ is an ordinary lattice field and $q^1$ is described by a $\Z$-gauge field $z^{(u)}$ (at each local trivialization labeled by $u\in U(1)$) could serve as a point of coincidence with \cite{AGGST19, NTU21, SG19, GGS18, SGG20, Sulej20}.

\section*{Acknowledgements}
We thank Mike Creutz and Wolfgang Bietenholz for advise and encouragement. We also thank Claudio Meneses, Juan Orendain and Manuel Sedano for discussions, and we thank Tin Sulejmanpasic for critically reading an earlier version of this manuscript. 
J. A. Zapata was supported in part by grant 
PAPIIT-UNAM IN100218. P.~Dall'Olio was supported by a DGAPA-UNAM grant.

\medskip

\bibliographystyle{unsrt}

\end{document}